\documentclass[12pt]{iopart}
\usepackage{iopams}
\usepackage{xcolor}
\usepackage{hyperref}
\usepackage{graphicx}
\usepackage{subfig}
\usepackage{booktabs}
\usepackage{multirow}
\usepackage{cite}

\begin{document}

\title[Three-dimensional boundary turbulence simulations of a RFX-mod plasma\dots]{Three-dimensional boundary turbulence simulations of a RFX-mod plasma in the presence of voltage biasing}

\author{M. Giacomin$^{1,2}$, N. Vianello$^{2,3}$, R. Cavazzana$^{2}$, S. Molisani$^{2,4}$, M. Spolaore$^{2,3}$ and M. Zuin$^{2,3}$}

\address{$^1$Dipartimento di Fisica ``G. Galilei'', Università degli Studi di Padova, 35121, Padua, Italy}
\address{$^2$Consorzio RFX, Corso Stati Uniti 4, 35127, Padua, Italy}
\address{$^3$Istituto per la Scienza e la Tecnologia dei Plasmi, CNR, Padua, Italy}
\address{$^4$Centro Ricerche Fusione, Università degli Studi di Padova, 35127, Padua, Italy. }

\ead{maurizio.giacomin@unipd.it}

\begin{abstract}
Three-dimensional turbulence simulations of a RFX-mod diverted plasma are performed in the presence of a biasing electrode. The simulations  show a strong suppression of turbulent transport caused by the induced $\mathbf{E}\times\mathbf{B}$ flow shear, which leads to the formation of an edge transport barrier with a pedestal-like structure, in qualitative agreement with RFX-mod experiments.
The strong $\mathbf{E}\times\mathbf{B}$ flow shear turbulence suppression with edge voltage biasing is also observed in the proximity of the density limit crossing, suggesting that edge voltage biasing may allow for larger maximum achievable density values.
By leveraging the simulation results, the theoretical scaling law of the edge pressure gradient length derived in Giacomin \& Ricci (2020) J. Plasma Phys. 86(5) is extended here to account for the $\mathbf{E}\times\mathbf{B}$ flow shear turbulence suppression caused by voltage biasing.
The improved theoretical scaling with typical RFX-mod shearing rate values predicts a factor of two increase of the pressure gradient at the separatrix, which is comparable to RFX-mod experiments in the presence of voltage biasing. The implications of the flow shear turbulence suppression due to voltage biasing on the density limit in RFX-mod are also discussed.          

\end{abstract}


\section{Introduction}

The $\mathbf{E}\times\mathbf{B}$ velocity shear plays a key role in suppressing turbulent transport in tokamaks, enabling the transition to a high-confinement regime (H-mode)~\cite{Wagner1982,biglari1990,Burrell1997,kim2003,ryter2013}.
A direct method to generate $\mathbf{E}\times\mathbf{B}$ flow shear consists in biasing the tokamak plasma boundary region. This is typically achieved by directly biasing the limiter or the divertor plates, or through a polarized electrode inserted into the plasma boundary. 
Since the pioneering work of Taylor \emph{et al}~\cite{taylor1989}, experiments carried out in DIII-D~\cite{shimada1990}, ISTTOK~\cite{cabral1998,silva2003}, RFX-mod~\cite{spolaore2017,grenfell2020} and TEXTOR~\cite{weynants1992,boedo2000} tokamaks have shown a clear reduction of turbulent transport when voltage biasing is applied at the tokamak boundary, often inducing a low-to-high confinement transition (L-H transition) with a heating power below the power threshold required for a spontaneous L-H transition.
A good correlation between confinement modifications and $\mathbf{E}\times\mathbf{B}$ flow shear has been found in experiments with boundary voltage biasing~\cite{silva2003}, suggesting that confinement enhancement originates at the plasma boundary as a result of reduced  turbulence.
Biasing experiments have also contributed significantly to the understanding of the role played by turbulence and flows in the L-H transition.
In particular, past RFX-mod experiments have shown that the H-mode can be routinely and robustly achieved at low power by inducing $\mathbf{E}\times\mathbf{B}$ flow shear by means of a biasing electrode inserted into the plasma edge of a diverted magnetic configuration~\cite{spolaore2017,grenfell2020}.  

Analogously, theoretical and numerical works have been carried out over the past years with the aim of investigating the effect of the $\mathbf{E}\times\mathbf{B}$ flow shear induced by voltage biasing on boundary turbulence~\cite{stringer1993,ghendrih2003,kasuya2003,shankar2022,shankar2024}. These works agree with the experimental observations and confirm that turbulence can be strongly suppressed by inducing $\mathbf{E}\times\mathbf{B}$ flow shear~\cite{stringer1993,benkadda2002,ghendrih2003,kasuya2003,shankar2022,shankar2024}, potentially allowing for the L-H transition.
The edge $\mathbf{E}\times\mathbf{B}$ velocity shear may also play an important role in regulating the maximum density that can be achieved in tokamak, denoted as the density limit, as suggested in a recent theoretical work~\cite{singh2022}, where the collapse of zonal flow shear has been proposed as the main mechanism behind the density limit~\cite{singh2022}. We note, however, that there is not a general consensus on the mechanisms setting the density limit, and various theories, which do not rely on flow shear, have been proposed in the past years~\cite{greenwald2002,gates2012,zanca2017,eich2021,giacomin2022density}.

Motivated by these previous experimental, theoretical and numerical results and in view of the upcoming RFX-mod2 experiments, we present here the results of the first three-dimensional global turbulence simulations carried out with the GBS code~\cite{Ricci2012,giacomin2022gbs} in a realistic RFX-mod diverted plasma in the presence of an edge biasing electrode. These simulations are performed at both low and high density with the aim of investigating the effect of voltage biasing near the L-H transition and the density limit.
By leveraging the GBS simulation results, the theoretical scaling law of the equilibrium pressure gradient length near the separatrix derived in~\cite{giacomin2020transp} and successfully validated against a multi-machine database in~\cite{giacomin2021} is extended here to include the  $\mathbf{E}\times\mathbf{B}$ flow shear turbulence suppression, which was previously neglected in~\cite{giacomin2020transp}.
Although this analysis is restricted to the resistive ballooning mode (RBM) regime, we highlight that flow shear turbulence suppression has been observed experimentally and numerically in regimes where turbulence is driven by different plasma microinstabilities, including ion temperature gradient~\cite{artun1992,hobbs2001}, kinetic ballooning modes~\cite{giacomin2024}, microtearing modes~\cite{guttenfelder2011,hardman2023,patel2024} and electron temperature gradient~\cite{roach2009,ivanov2024}. Therefore, some of the results presented here from turbulence simulations in the RBM regime with voltage biasing may hold in other turbulent transport regimes.
  
The paper is organized as follows. The physical model, the numerical implementation of the electrode and an overview of GBS simulation results are presented in section~\ref{sec:gbs}. The  dependence of the equilibrium pressure gradient on the $\mathbf{E}\times\mathbf{B}$ velocity shearing rate is analyzed in section~\ref{sec:nonlocal} by leveraging the results of non-local linear simulations. In section~\ref{sec:theory}, the results of linear and nonlinear simulations are exploited to derive an analytical estimate of the $\mathbf{E}\times\mathbf{B}$ flow shear turbulence suppression factor, which is also used to address the impact of edge voltage biasing on the density limit. Conclusions follow in section~\ref{sec:conclusions}. 

\section{GBS simulations of a RFX-mod plasma with edge biasing}
\label{sec:gbs}

This section investigates the effect of edge negative voltage biasing on boundary turbulence in RFX-mod plasmas by leveraging the results of flux-driven global two-fluid turbulence simulations carried out with the GBS code. 
GBS has been extensively used in the past to perform boundary turbulence simulations in complex magnetic configurations, including snowflake configurations~\cite{giacomin2020snow} and stellarator geometries~\cite{coelho2022}, and it has been validated against a number of dedicated experiments (see, e.g., \cite{galassi2022,oliveira2022}). Therefore, GBS is a suitable code to perform three-dimensional flux-driven turbulence simulations with edge voltage biasing in a realistic diverted plasma geometry, as described in the following. 

\subsection{Physical model and numerical implementation of the electrode in GBS}

The physical model considered in the present work is based on the drift-reduced Braginskii model~\cite{Zeiler1997} implemented in GBS~\cite{giacomin2022gbs}. Given the low values of $\beta$ in RFX-mod tokamak discharges, electromagnetic effects are expected to play a secondary role, hence the electrostatic limit of GBS is considered here. The neutral dynamics is neglected here, since the plasma-neutral interaction has been found to weakly affect the equilibrium pressure gradient close to the separatrix in the recycling condition typically explored on RFX-mod~\cite{mancini2021}, which is the main focus of the present work.

With these approximations, the model equations are
\begin{eqnarray}
\label{eqn:density}
\fl\frac{\partial n}{\partial t} &=& -\frac{\rho_*^{-1}}{B}\bigl[\phi,n\bigr]+\frac{2}{B}\Bigl[C(p_e)-nC(\phi)\Bigr] 
-\nabla_{\parallel}(n v_{\parallel e}) + D_n\nabla_{\perp}^2 n +s_n\, ,\\
\label{eqn:vorticity}
\fl\frac{\partial \Omega}{\partial t} &=& -\frac{\rho_*^{-1}}{B}\nabla \cdot [\phi,\boldsymbol{\omega}] - \nabla \cdot \bigl( v_{\parallel i}\nabla_\parallel \boldsymbol{\omega}\bigr) + B^2\nabla_{\parallel}j_{\parallel} + 2B C(p_e + \tau p_i) \nonumber\\
\fl&+& \frac{B}{3}C(G_i) + D_{\Omega}\nabla_\perp^2 \Omega\,,\\
\label{eqn:electron_velocity}
\fl\frac{\partial v_{\parallel e}}{\partial t} &=& -\frac{\rho_*^{-1}}{B}\bigl[\phi,v_{\parallel e}\bigr] - v_{\parallel e}\nabla_\parallel v_{\parallel e} 
+ \frac{m_i}{m_e}\Bigl(\nu j_\parallel+\nabla_\parallel\phi-\frac{1}{n}\nabla_\parallel p_e-0.71\nabla_\parallel T_e\Bigr) \nonumber \\
\fl&+&\frac{4}{3n}\frac{m_i}{m_e}\eta_{0,e}\nabla^2_\parallel v_{\parallel e} + D_{v_{\parallel e}}\nabla_\perp^2 v_{\parallel e}\,, \\
\label{eqn:ion_velocity}
\fl\frac{\partial v_{\parallel i}}{\partial t} &=& -\frac{\rho_*^{-1}}{B}\bigl[\phi,v_{\parallel i}\bigr] - v_{\parallel i}\nabla_\parallel v_{\parallel i} - \frac{1}{n}\nabla_\parallel(p_e+\tau p_i)
+ \frac{4}{3n}\eta_{0,i}\nabla^2_\parallel v_{\parallel i} + D_{v_{\parallel i}}\nabla_\perp^2 v_{\parallel i}\, ,\\
\label{eqn:electron_temperature}
\fl\frac{\partial T_e}{\partial t} &=& -\frac{\rho_*^{-1}}{B}\bigl[\phi,T_e\bigr]  
+ \frac{2}{3}T_e\Bigl[0.71\nabla_\parallel v_{\parallel i}-1.71\nabla_\parallel v_{\parallel e}
+0.71 (v_{\parallel i}-v_{\parallel e})\frac{\nabla_\parallel n}{n}\Bigr] \nonumber \\
\fl&-& v_{\parallel e}\nabla_\parallel T_e + \frac{4}{3}\frac{T_e}{B}\Bigl[\frac{7}{2}C(T_e)+\frac{T_e}{n}C(n)-C(\phi)\Bigr] 
+ \chi_{\parallel e}\nabla_\parallel^2 T_e - 2.61 \nu n (T_e - T_i)\nonumber\\
\fl&+& D_{T_e}\nabla_\perp^2 T_e + s_{T_e}\,,\\
\label{eqn:ion_temperature}
\fl\frac{\partial T_i}{\partial t} &=& -\frac{\rho_*^{-1}}{B}\bigl[\phi,T_i\bigr] - v_{\parallel i}\nabla_\parallel T_i 
+ \frac{4}{3}\frac{T_i}{B}\Bigl[C(T_e)+\frac{T_e}{n}C(n)-C(\phi)\Bigr] - \frac{10}{3}\tau\frac{T_i}{B}C(T_i)\nonumber\\
\fl&-&\frac{2}{3}T_i\nabla_\parallel v_{\parallel e} + \frac{2}{3}T_i(v_{\parallel i}-v_{\parallel e})\frac{\nabla_\parallel n}{n}  
+ \chi_{\parallel i}\nabla_\parallel^2 T_i + 2.61 \nu n (T_e - T_i)\nonumber\\
\fl&+&   D_{T_i}\nabla_\perp^2 T_i + s_{T_i}\,,\\
\label{eqn:poisson}
\fl\nabla \cdot \bigl( n&\nabla_\perp& \phi\bigr) = \ \Omega-\tau\nabla_\perp^2 p_i\,,
\end{eqnarray}
where $n$ is the density, $\Omega = \nabla\cdot\boldsymbol{\omega} = \nabla \cdot (n \nabla_\perp\phi + \tau \nabla_\perp p_i)$ is the scalar vorticity, $v_{\parallel e}$ and $v_{\parallel i}$ are the electron and ion parallel velocities, $T_e$ and $T_i$ are the electron and ion temperatures, $p_e = nT_e$ and $p_i=nT_i$ are the electron and ion pressures, and $\phi$ is the electrostatic potential.  

In equations~(\ref{eqn:density})-(\ref{eqn:poisson}) and in the following, we use GBS normalized units, where $n$, $T_e$ and $T_i$ are normalized to the reference values $n_0$, $T_{e0}$ and $T_{i0}$, respectively, $v_{\parallel e}$ and $v_{\parallel i}$, are normalized to the reference sound speed $c_{s0}=\sqrt{T_{e0}/m_i}$, $\phi$ is normalized to $T_{e0}/e$, the magnetic field is normalized to $B_0$, lengths perpendicular to the magnetic field are normalized to $\rho_{s0}= c_{s0}/\Omega_{ci}$, with $\Omega_{ci} = e B_0/m_i$ the ion cyclotron frequency, lengths parallel to the magnetic field are normalized to the tokamak major radius $R_0$, and time is normalized to $R_0/c_{s0}$. 
The following dimensionless parameters appear in equations~(\ref{eqn:density})-(\ref{eqn:poisson}):  $\rho_* = \rho_{s0}/R_0$, $\tau = T_{i0}/T_{e0}$, the normalized electron and ion parallel thermal conductivities,
\begin{eqnarray}
\label{eqn:chie}
    \chi_{\parallel e} &=& \chi_{\parallel e 0}T_e^{5/2} = \biggl(\frac{1.58}{\sqrt{2\pi}} \frac{m_i}{\sqrt{m_e}} \frac{(4\pi\epsilon_0)^2}{e^4} \frac{c_{s0}}{R_0} \frac{T_{e0}^{3/2}}{\lambda n_0}\biggr) T_e^{5/2}\,,\\
    \chi_{\parallel i} &=& \chi_{\parallel i 0}T_i^{5/2} = \biggl(\frac{1.94}{\sqrt{2\pi}} \sqrt{m_i} \frac{(4\pi\epsilon_0)^2}{e^4} \frac{c_{s0}}{R_0} \frac{T_{e0}^{3/2}\tau^{5/2}}{\lambda n_0}\biggr) T_i^{5/2}\,,
\end{eqnarray}
and the normalized Spitzer resistivity, $\nu = e^2n_0R_0/(m_ic_{s0}\sigma_\parallel) = \nu_0 T_e^{-3/2}$, where
\begin{eqnarray}
\label{eqn:conductivity}
\sigma_\parallel &=& \biggl(1.96\frac{n_0 e^2 \tau_e}{m_e}\biggr)n=\biggl(\frac{5.88}{4\sqrt{2\pi}}\frac{(4\pi\epsilon_0)^2}{e^2}\frac{ T_{e0}^{3/2}}{\lambda\sqrt{m_e}}\biggr)T_e^{3/2},\\
\label{eqn:resistivity}
\nu_0 &=&\frac{4\sqrt{2\pi}}{5.88}\frac{e^4}{(4\pi\epsilon_0)^2}\frac{\sqrt{m_e}R_0n_0\lambda}{m_i c_{s0} T_{e0}^{3/2}},
\end{eqnarray}
with $\lambda$ the Coulomb logarithm.
Further details on the physical model, including the definition of the gyroviscous terms $G_e$ and $G_i$, are reported in~\cite{giacomin2022gbs}.

The spatial differential operators appearing in equations~(\ref{eqn:density})-(\ref{eqn:poisson}) are the $\mathbf{E}\times\mathbf{B}$ convective term $\bigl[\phi,f\bigr]=\mathbf{b}\ \cdot\ (\nabla \phi \times \nabla f)$, the curvature operator $C(f)=B\bigl[\nabla \times (\mathbf{b}/B)\bigr]/2\cdot \nabla f$, the perpendicular Laplacian operator ${\nabla_\perp^2 f=\nabla\cdot\bigl[(\mathbf{b}\times\nabla f)\times\mathbf{b}\bigr]}$ and the parallel gradient operator $\nabla_\parallel f=\mathbf{b}\cdot\nabla f$, where $\mathbf{b}=\mathbf{B}/B$ is the unit vector of the magnetic field. 
These operators are discretized on a uniform cylindrical grid $(R, \varphi, Z)$.
The source terms $s_n$ and $s_T$ are toroidally symmetric functions defined as 
\begin{eqnarray}
    \label{eqn:density_source}
    s_n &=& s_{n0} \exp\biggl(-\frac{\bigl(\Psi(R,Z)-\Psi_{n}\bigr)^2}{\Delta_n^2}\biggr),\\   
    \label{eqn:temperature_source}
    s_T &=& \frac{s_{T0}}{2}\biggl[\tanh\biggl(-\frac{\Psi(R,Z)-\Psi_{T}}{\Delta_T}\biggr)+1\biggr], 
\end{eqnarray}
where $\Psi_n$ and $\Psi_T$ are flux surfaces located inside the last-closed flux surface ($A_{\mathrm{LCFS}}$). Similarly to~\cite{giacomin2020transp}, we define the total density and temperature source integrated over the area inside the last-closed flux-surface, i.e.
\begin{equation}
    S_n=\int_{A_{\mathrm{LCFS}}} \rho_* s_n(R,Z)\,\mathrm{d}R\mathrm{d}Z
\end{equation}
and
\begin{equation}
    S_T=\int_{A_{\mathrm{LCFS}}} \rho_* s_T(R,Z)\,\mathrm{d}R\mathrm{d}Z\,,
\end{equation}
as well as the total electron pressure source,  $S_p=\int_{A_{\mathrm{LCFS}}} \rho_* s_p\,\mathrm{d}R\mathrm{d}Z$, with $s_p=n s_{T_e} + T_e s_n$.
Details on the numerical implementation of equations~(\ref{eqn:density})-(\ref{eqn:poisson}) in GBS are reported in~\cite{giacomin2022gbs}. 

\subsubsection{Implementation of the biasing electrode in GBS}
\hfill\\
The biasing electrode is implemented in GBS through the pre-sheath boundary conditions of Poisson's equation (see equation~(\ref{eqn:poisson})) in correspondence of the electrode head, in analogy with the implementation of the wall boundary conditions detailed in~\cite{giacomin2022gbs}. The electrostatic potential in the plasma region occupied by the electrode is therefore imposed as a finite Dirichlet boundary condition, thus mimicking the experimental setup of RFX-mod\footnote{Technical details on the RFX-mod electrode, including details on the electrode surface heating and on the electrode power supply, are reported in~\cite{spolaore2017}.}. The value of the electrostatic potential, the position and the size of the electrode can be freely chosen to match the experimental configuration.
Neumann zero boundary conditions are considered at the magnetic pre-sheath entrance of the plasma-electrode interface for all the quantities other than the electrostatic potential.
This set of boundary conditions at the electrode head retains the primary effect of the biasing electrode in modifying the electrostatic potential and, thus, inducing an $\mathbf{E}\times\mathbf{B}$ flow shear, while the local sputtering and recycling at the electrode head are neglected.
We highlight that the chosen boundary conditions at the electrode head allow for particles, heat and current to flow through the electrode.
Furthermore, the current implementation of the electrode is limited to axisymmetric magnetic configurations, with an electrode that is localized poloidally and extends in the toroidal direction over the full torus.

\subsection{Overview of the RFX-mod simulations}

We consider here a set of GBS simulations carried out with the RFX-mod magnetic equilibrium of the discharge \#39136 ($a=0.46$~m, $R_0=2$~m, $B_0=0.55$~T). The reference quantities are taken at the separatrix of the chosen RFX-mod discharge ($n_0\simeq 2\cdot 10^{18}$~m$^{-3}$, $T_{e0}\simeq 20$~eV).
The reference ion sound Larmor radius is computed from the reference temperature and magnetic field and is approximately $\rho_{s0}\simeq 1.2$~mm. In order to reduce the computational cost of the simulation scan, we perform simulations with a domain corresponding to half-size of RFX-mod, i.e. $L_R = 450\;\rho_{s0}$, $L_Z = 420\;\rho_{s0}$ and $\rho_{*}^{-1} = R_0/\rho_{s0} = 850$, deferring full-size RFX-mod simulations to a dedicated future validation.
The number of grid points in the radial, toroidal and vertical direction is $(N_R, N_\varphi, N_Z) = (200, 64, 200)$. The simulation time step is $\Delta t = 4\times 10^{-6}\, R_0/c_{s0}$.
Given the reduced simulation domain with respect to RFX-mod, the density and temperature sources (see equations~(\ref{eqn:density_source})~and~(\ref{eqn:temperature_source})) are chosen such that the value of the total fueling rate and heating power is approximately a factor of four smaller than the experimental value. In particular, we choose $s_{n0} = 0.05$, $s_{T_{e}0} = 0.05$ and $s_{T_{i}0} = 0$ (there is no external ion heating in RFX-mod).
Two different values of $\nu_0$ (which is proportional to the reference density, see equation~(\ref{eqn:resistivity})) are considered here, $\nu_0=0.1$ and $\nu_0=1.0$. The values of $\nu_0$ are chosen according to the phase space of edge turbulence derived in~\cite{giacomin2022turbulent}, such that the low $\nu_0$ simulation is near the transition between the RBM and the drift-wave regimes, and the high $\nu_0$ simulation is above the crossing of the density limit.
The low $\nu_0$ value is comparable to the experimental one near the separatrix of the reference discharge. On the other hand, the high $\nu_0$ case corresponds to a reference density that is an order of magnitude larger than the separatrix density of the discharge \#39136.
The GBS simulations at low and high density are performed with and without edge voltage biasing, leading to a total of four simulations.
In both the biased simulations, the electrode is located in the bottom of the domain, just inside the separatrix at approximately $r/a \simeq 0.9$, thus matching the experimental configuration of the RFX-mod device. The tokamak edge is biased to negative voltage values (with respect to the wall) that are similar to those applied in the RFX-mod discharge \#39136.
Simulations are performed until a quasi-steady state is reached. Equilibrium quantities are computed as time and toroidal average in the quasi-steady state phase of the simulations. 

\begin{figure}
    \centering
    \subfloat[$\nu_0=0.1$, unbiased]{\includegraphics[width=0.24\linewidth]{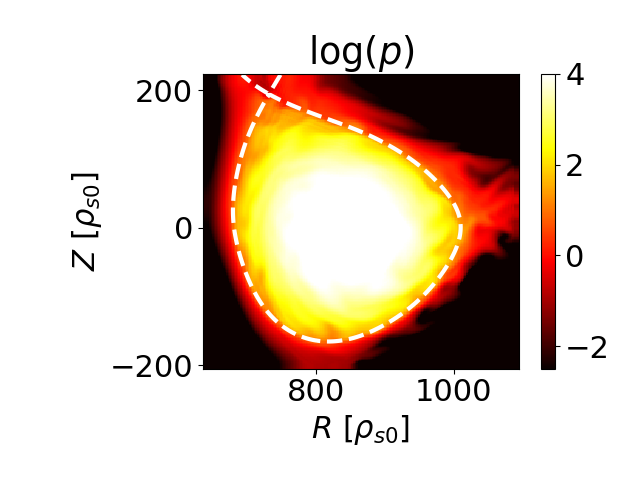}}
    \subfloat[$\nu_0=0.1$, biased]{\includegraphics[width=0.24\linewidth]{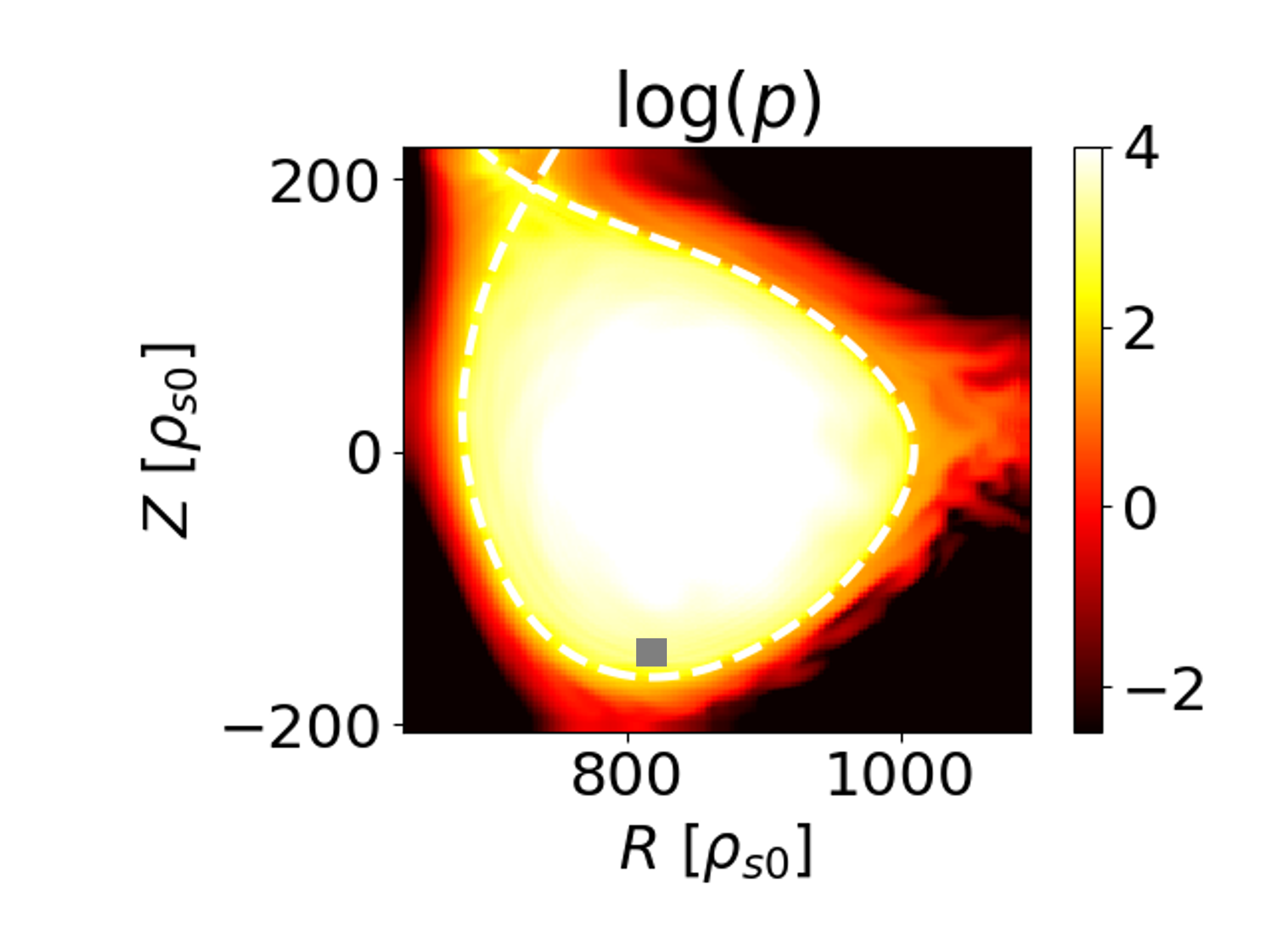}}
    \subfloat[$\nu_0=1.0$, unbiased]{\includegraphics[width=0.24\linewidth]{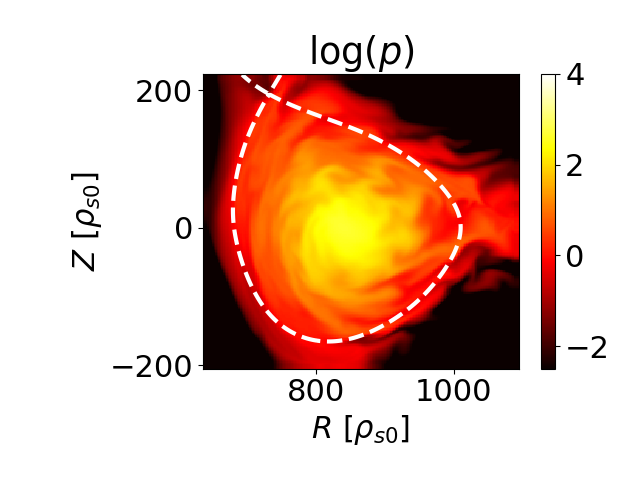}}
    \subfloat[$\nu_0=1.0$, biased]{\includegraphics[width=0.24\linewidth]{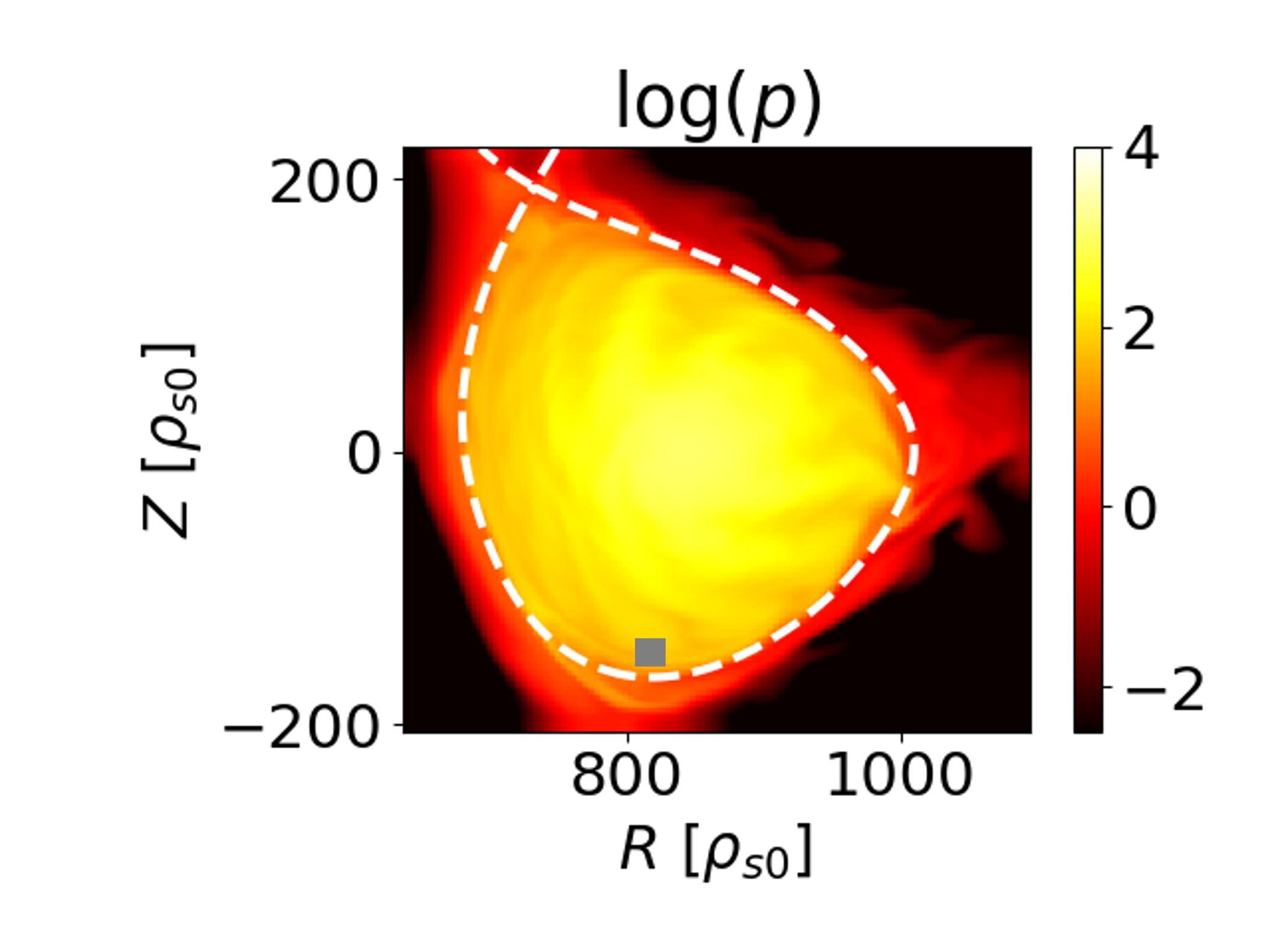}}\\
    \subfloat[$\nu_0=0.1$, unbiased]{\includegraphics[width=0.24\linewidth]{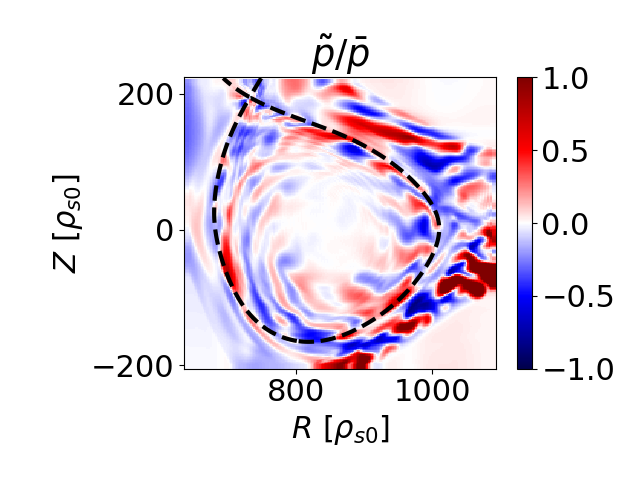}}
    \subfloat[$\nu_0=0.1$, biased]{\includegraphics[width=0.24\linewidth]{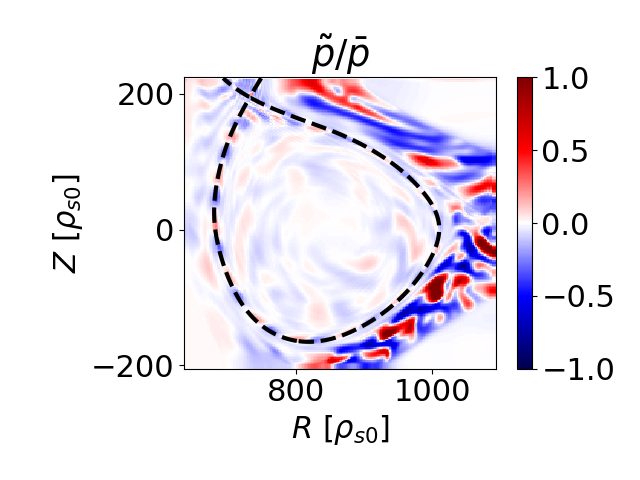}}
    \subfloat[$\nu_0=1.0$, unbiased]{\includegraphics[width=0.24\linewidth]{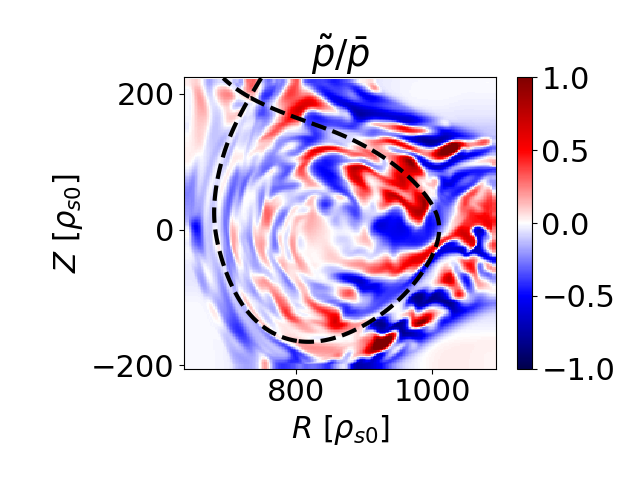}}
    \subfloat[$\nu_0=1.0$, biased]{\includegraphics[width=0.24\linewidth]{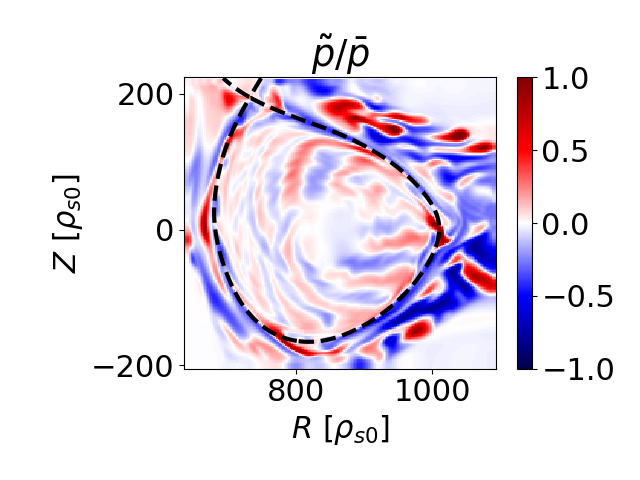}}
\caption{Two-dimensional snapshot of the pressure [(a)-(d)] and the corresponding relative fluctuations [(e)-(h)] from the four GBS simulations considered in this work. The dashed line denotes the separatrix location. The grey square in (b) and (d) indicates the location of the biasing electrode (not in scale for clarity purpose).}
\label{fig:pressure}
\end{figure}

Figure~\ref{fig:pressure} shows two-dimensional snapshots of the pressure and its fluctuations from the four GBS simulations. In the simulations without voltage biasing, the pressure confinement is heavily degraded at large $\nu_0$, in agreement with previous numerical investigations~\cite{giacomin2020transp} that show a regime of catastrophically large turbulent transport at high $\nu_0$. The size of the turbulent eddies increases with $\nu_0$ and the pressure fluctuations extend into the confined plasma region, which is very similar to the dynamics pointed out in~\cite{giacomin2020transp}, despite the different magnetic configuration considered here. 
A substantial reduction of the pressure fluctuation amplitudes is observed in the simulations with edge voltage biasing at both values of $\nu_0$, both in the tokamak edge, i.e. the region just inside the last-closed flux surface, and in the near scrape-off layer (SOL), i.e. the region near the last-closed flux surface where the magnetic field lines intercept the vessel wall. In particular, turbulence suppression is clearly visible at $\nu_0=0.1$ in the region across the separatrix (see figure~\ref{fig:pressure}(f)).
The amplitude of the pressure fluctuations and the size of the turbulent eddies are also reduced in the far SOL of the $\nu_0=1.0$ biased simulation, while the far SOL turbulence is less affected by the voltage biasing in the $\nu_0=0.1$ case, where the values of the equilibrium pressure and fluctuation amplitudes in the biased and unbiased case are similar.
It is worth noting that the flat pressure profile caused by large turbulent transport in the high-$\nu_0$ unbiased simulation is completely avoided in the biased simulation, although these two simulations share the same value of the model parameters (in particular heating source and $\nu_0$) and should belong, therefore, to the same turbulent transport regime of~\cite{giacomin2022turbulent}. 
However, the reduction of turbulent transport across the separatrix due to edge voltage biasing prevents the edge pressure gradient from collapsing, thus allowing the simulation to maintain good plasma confinement at values of $\nu_0$ larger than in the unbiased case, thus pointing out a potentially important effect of voltage biasing on the density limit, which is discussed in section~\ref{sec:density_limit}.

\begin{figure}
    \centering
    \subfloat[]{\includegraphics[width=0.48\linewidth]{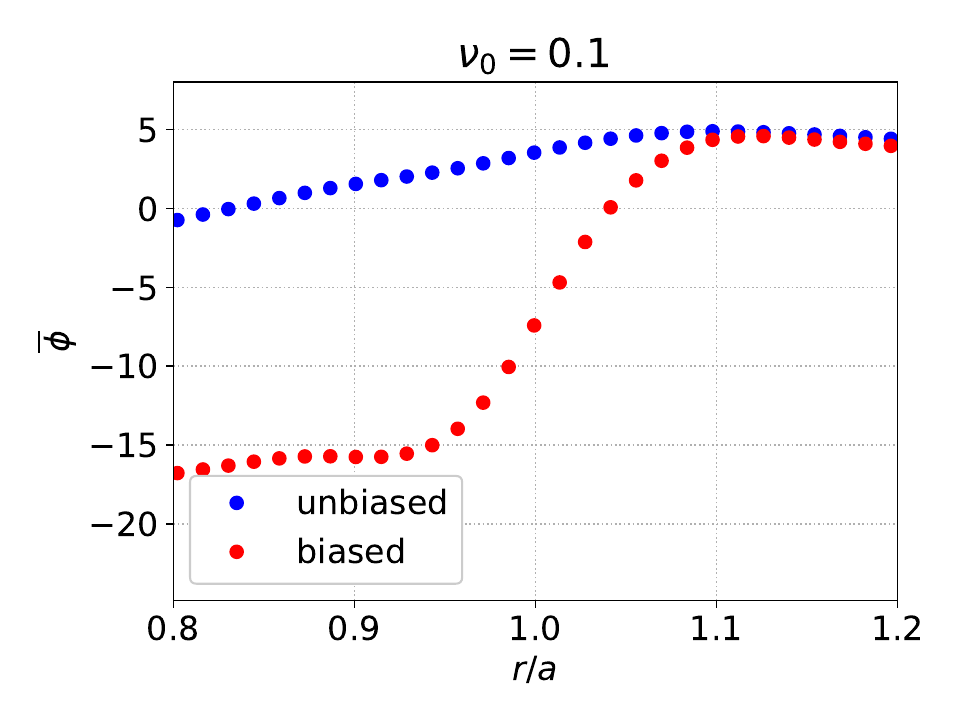}}
    \subfloat[]{\includegraphics[width=0.48\linewidth]{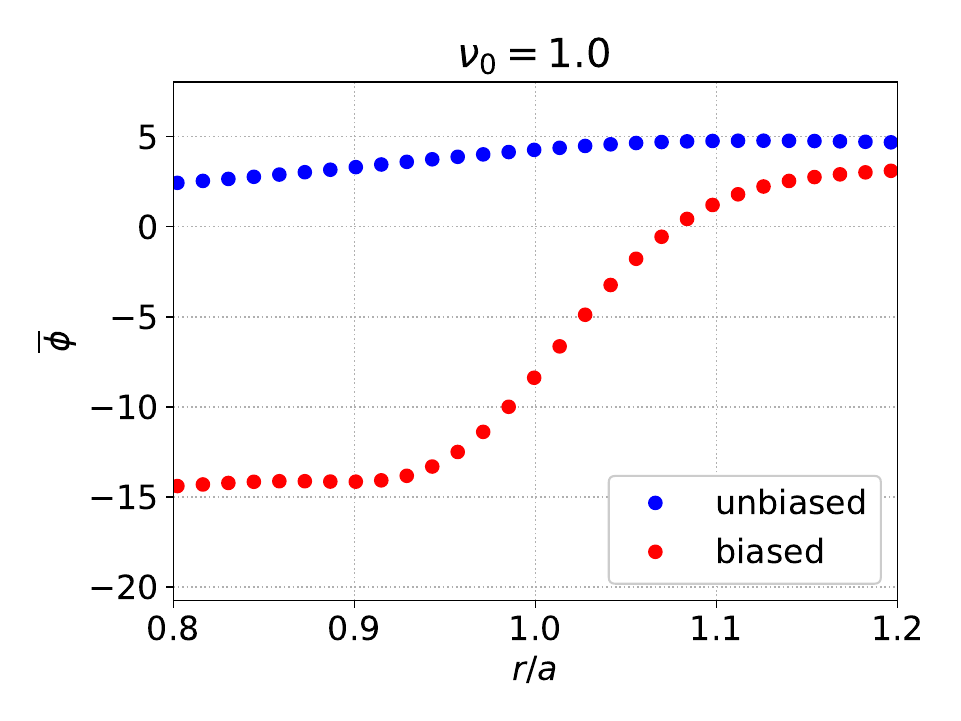}}\\
    \subfloat[]{\includegraphics[width=0.48\linewidth]{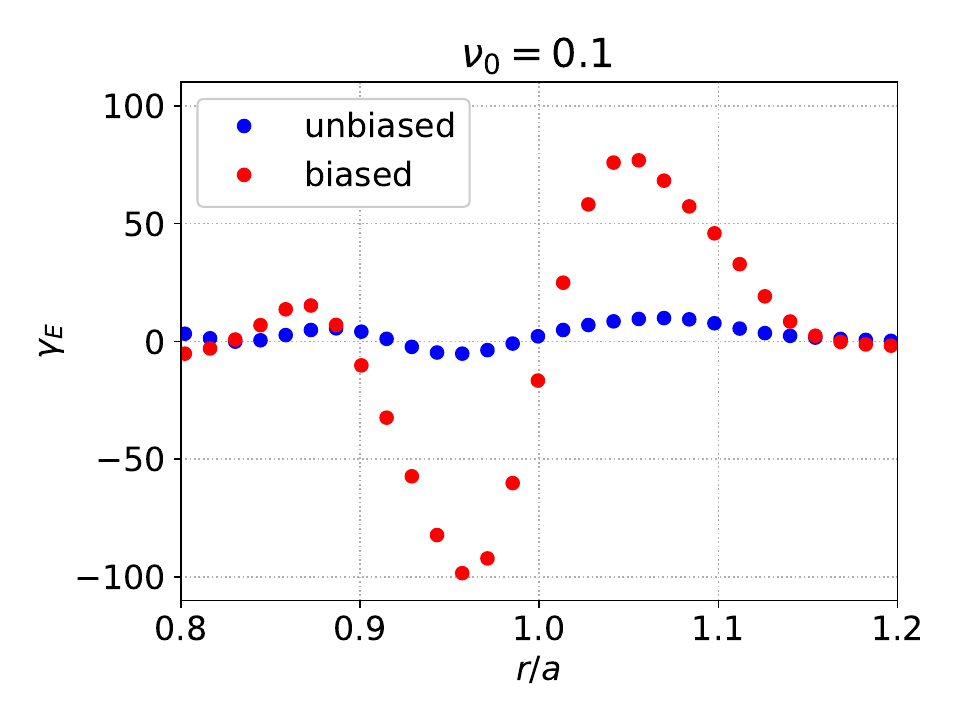}}
    \subfloat[]{\includegraphics[width=0.48\linewidth]{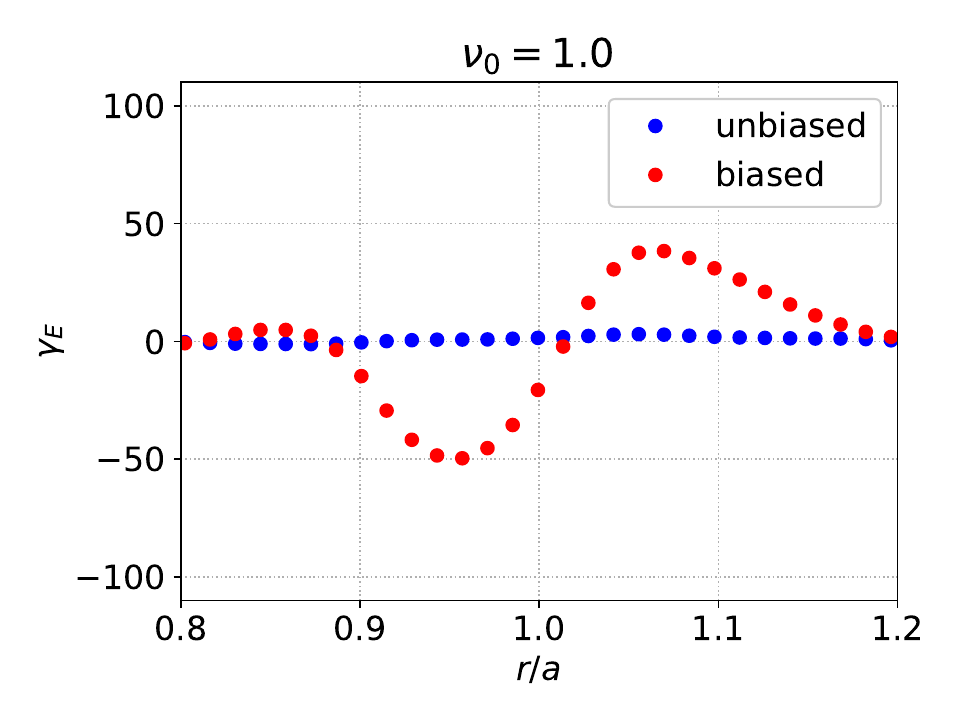}}
    \caption{Outboard mid-plane equilibrium electrostatic potential (top row) and $\mathbf{E}\times\mathbf{B}$ flow shear (bottom row) radial profiles in the proximity of the separatrix from the GBS simulations at $\nu_0=0.1$ [(a) and (c)] and $\nu_0=1.0$ [(b) and (d)] with (red markers) and without (blue markers) edge voltage biasing.}
    \label{fig:shear_profiles}
\end{figure}

Turbulence suppression in GBS simulations with edge voltage biasing is caused by the strongly induced $\mathbf{E}\times\mathbf{B}$ flow shear. Figure~\ref{fig:shear_profiles} compares the outboard mid-plane equilibrium electrostatic potential and $\mathbf{E}\times\mathbf{B}$ flow shear (defined as $\gamma_E = -\rho_*^{-1} \partial_{rr}\bar{\phi}$  and normalized to $c_{s0}/R_0$) radial profiles in the proximity of the separatrix as functions of the minor radius coordinate $r$ (normalized to the minor radius $a$) from the four GBS simulations.
The biasing electrode forces the electrostatic potential in the region near $r/a\simeq 0.9$ to a value close to $\phi_b\simeq -15$, corresponding to a biasing voltage, in physical units, of $V_b =T_{e0}\phi_b/e \simeq -300$~V, which is close to typical values achieved experimentally in RFX-mod~\cite{spolaore2017,grenfell2020}.
In the GBS simulations without voltage biasing, the major contribution to the edge electric field comes from the radial force balance, leading to an electric field that is approximately proportional to the ion pressure gradient, while zonal flow contributions generated by plasma turbulence are typically small. 
The $\mathbf{E}\times\mathbf{B}$ shearing rate is relatively weak in the unbiased simulations, especially at $\nu_0=1.0$, where the $\mathbf{E}\times\mathbf{B}$ flow shear is completely negligible. Consequently, the self-generated $\mathbf{E}\times\mathbf{B}$ flow shear turbulence suppression, which includes zonal and diamagnetic flows, is weak in the unbiased simulations, in agreement with previous simulation results~\cite{giacomin2022turbulent}.
However, when the electrode is biased, a large $\mathbf{E}\times\mathbf{B}$ flow shear is generated in the proximity of the separatrix, leading to shearing rate values that overcome largely the ones reached in the unbiased simulations and, consequently, to turbulence suppression.  
Moreover, although the electrode is located inside the separatrix, a modification of the electrostatic potential is also observed in the near SOL. On the other hand, the electrostatic potential in the far SOL is only weakly affected by the biasing electrode.
We note that the maximum shearing rate value is larger at low $\nu_0$ than at high $\nu_0$ despite the applied biasing voltage being the same. This difference is caused by a lower electrostatic potential in the SOL at large $\nu_0$.      
In the low-$\nu_0$ simulation, the maximum shearing rate (in absolute value) is achieved at $r/a\simeq 0.95$, where it reaches values, in physical units, of the order of $\max|\gamma_E c_{s0}/R_0| \simeq 10^{6}$~s$^{-1}$, which are comparable to those reached in the RFX-mod discharge \#39136.
Moreover, the $\mathbf{E}\times\mathbf{B}$ flow shear profile in the GBS simulation at $\nu_0=0.1$ agrees qualitatively and quantitatively with the one measured in the H-mode phase of \#39136 (see \ref{sec:comparison} for a comparison between the experimental and numerical flow shear radial profiles), although a detailed comparison between simulation results and experimental measurements would require full-size RFX-mod simulations and it is postponed to a future dedicated validation work.

\begin{figure}
    \centering
    \subfloat[Unbiased]{\includegraphics[width=0.47\linewidth]{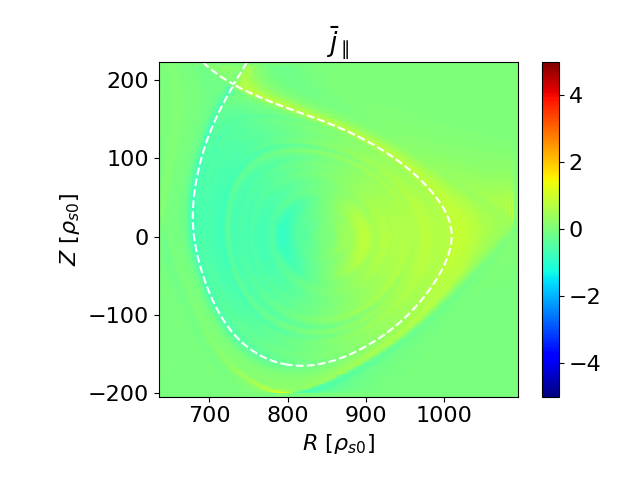}}
    \subfloat[Biased]{\includegraphics[width=0.47\linewidth]{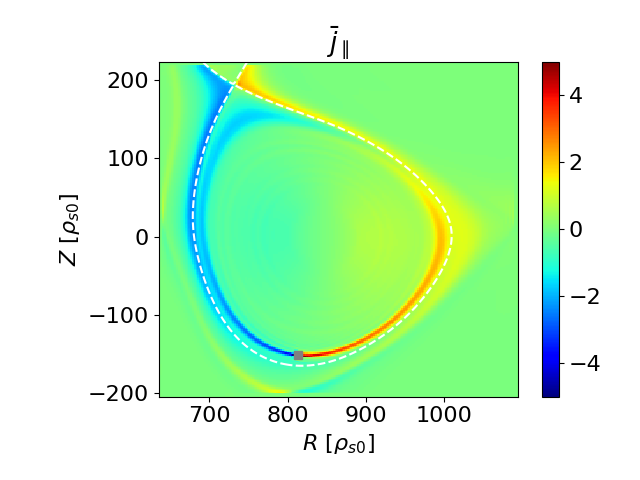}}
\caption{Equilibrium parallel current from the GBS simulations at $\nu_0=1.0$ without (a) and with (b) edge voltage biasing. The electrode position is indicated in (b) by the grey marker. The dashed line denotes the separatrix location.}
\label{fig:current}
\end{figure}

The presence of the biasing electrode gives rise to parallel currents flowing in the tokamak edge towards the electrode head. As shown in figure~\ref{fig:current}, these parallel currents are very large in the simulation at $\nu_0=1.0$ (parallel currents are significantly smaller in the GBS simulation with $\nu_0=0.1$). 
At high $\nu_0$, the total current drained by the electrode at a biasing voltage similar to the one in \#39136 exceeds 1~kA, which is an order of magnitude larger than the typical currents drained by the electrode in RFX-mod experiments~\cite{spolaore2017}. The large current drained by the electrode in the high-$\nu_0$  simulation (with a voltage biasing value near the experimental one) is a consequence of the high (reference) density, which is an order of magnitude larger than the separatrix density value in the RFX-mod discharge \#39136.
The substantial turbulent transport reduction at high $\nu_0$ with voltage biasing comes at the cost of a large current drained by the electrode: reaching these large current values may not be possible experimentally, considering also that the heat and particle fluxes flowing to the electrode head could lead to excessive sputtering or local surface melting, with detrimental effects for plasma confinement.
However, experimentally relevant values of the $\mathbf{E}\times\mathbf{B}$ flow shear at large density, able to suppress plasma turbulence, may be obtained with a lower biasing (and therefore lower drained current) compared to the low density case. 
This important topic will be addressed in future RFX-mod2 experiments at high density in the presence of voltage biasing.

\begin{figure}
    \centering
    \subfloat[]{\includegraphics[width=0.48\linewidth]{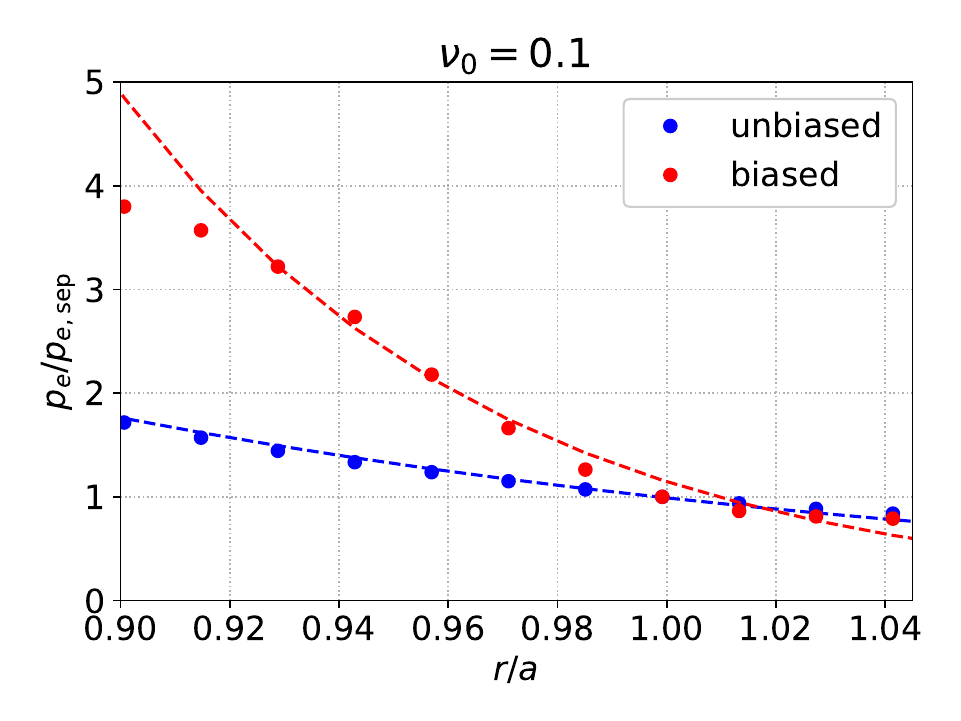}}
    \subfloat[]{\includegraphics[width=0.48\linewidth]{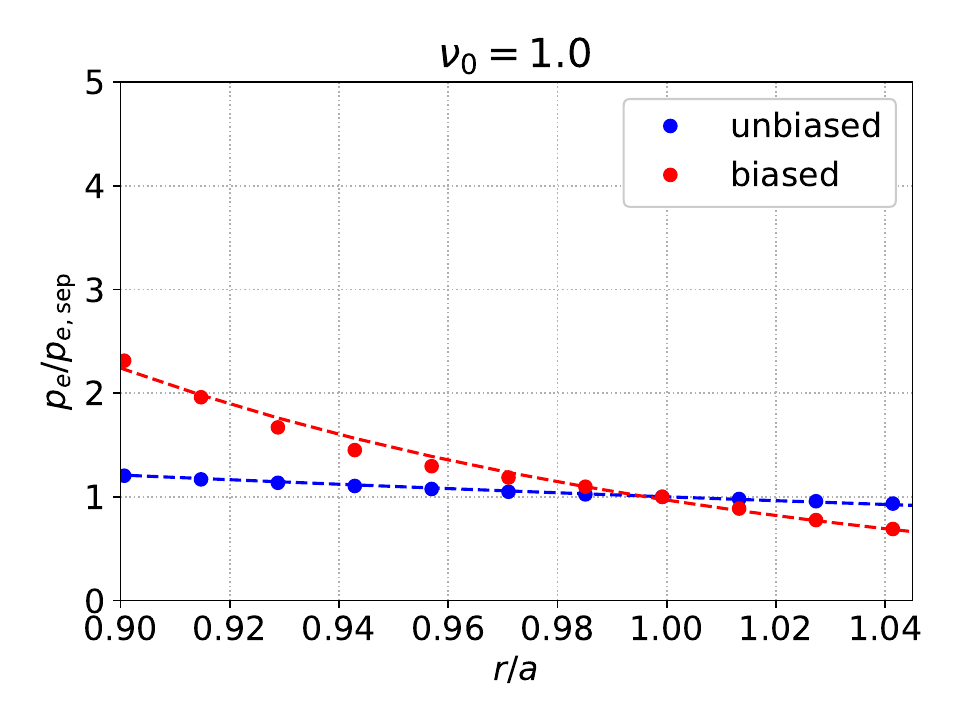}}
    \caption{Outboard mid-plane equilibrium electron pressure profiles in the proximity of the separatrix from the GBS simulations at $\nu_0=0.1$ (a) and $\nu_0=1.0$ (b) with (red markers) and without (blue markers) edge voltage biasing. The dashed lines show the exponential fit performed on the pressure profiles to evaluate $L_p$.}
    \label{fig:pressure_profiles}
\end{figure}

Figure~\ref{fig:pressure_profiles} compares the outboard mid-plane equilibrium electron pressure profiles in the proximity of the separatrix from the GBS simulations at $\nu_0=0.1$ and $\nu_0=1.0$ with and without edge voltage biasing. The induced $\mathbf{E}\times\mathbf{B}$ flow shear turbulence suppression leads to a steepening of the pressure profile in the biased simulations.
As discussed above, the difference on the pressure profile between the unbiased and biased case is remarkable at high $\nu_0$, with the pressure gradient of the high-$\nu_0$ biased simulation being comparable to the one of the low-$\nu_0$ unbiased simulation. 
The largest pressure gradient is reached in the biased simulation at $\nu_0=0.1$, where a pedestal-like transport barrier appears in the tokamak edge. The formation of a pedestal in the low-$\nu_0$ biased simulation resembles qualitatively the H-mode transition observed in RFX-mod experiments when the tokamak edge is biased to negative voltage values~\cite{spolaore2017,grenfell2020}.

The equilibrium pressure gradient lengths of the four simulations are evaluated by fitting the outboard mid-plane pressure profiles in figure~\ref{fig:pressure_profiles} with the exponential function $p_e = p_{e,\mathrm{sep}}\exp(-(r-r_\mathrm{sep})/L_p)$, depicted in figure~\ref{fig:pressure_profiles} as a dashed line. The numerical values of $L_p$ obtained from the fit are reported in table~\ref{tab:lp}. The value of $L_p$ in the biased case is approximately a factor of two and a factor of five smaller than the unbiased case at $\nu_0=0.1$ and $\nu_0=1.0$, respectively. 
The $\mathbf{E}\times\mathbf{B}$ flow shear turbulence suppression is therefore more effective at large $\nu_0$,  pointing out a non-trivial dependence of $L_p$ on $\gamma_E$, which  involves at least an extra dependence on $\nu_0$, but it is likely to also include dependencies on other parameters, such as the heating source and the safety factor. The dependence of $L_p$ on $\gamma_E$ is investigated in section~\ref{sec:nonlocal}.

In table~\ref{tab:lp}, the numerical values of $L_p$ obtained from the fitting are compared to the theoretical predictions derived in~\cite{giacomin2020transp},
\begin{equation}
\label{eqn:lp0}
    L_{p0} \simeq \biggl[\frac{\rho_*}{2}(\nu q^2 n)^2 \biggl(\frac{L_\chi}{S_p}p_e\biggr)^4\biggr]^{1/3}T_e \,,
\end{equation}
where $n$, $T_e$ and $p_e$ are the equilibrium density, electron temperature and electron pressure evaluated at the separatrix\footnote{We remind that the quantities $n$ and $T_e$ are normalized to the reference values $n_0$ and $T_{e0}$, which are taken near the separatrix. The value of $n$ and $T_e$ at the separatrix should be, therefore, close to unity. However, this value is not forced to be exactly one, since the kinetic profiles evolve self-consistently with sources and transport in GBS simulations. For example, in the GBS simulations considered here, $n\simeq 1$ and $T_e\simeq 1.5$ at the separatrix.}, $q\simeq 3.5$ is the safety factor at the surfaces encompassing the 95\% of the poloidal flux (also refereed to as $q_{95}$),  $S_p\simeq 20$ is the total heating source, which is approximately constant in the four simulations, and $L_\chi\simeq 1000$ is the length of the last-closed flux surface on the poloidal plane. 
As a side note, we mention that equation~(\ref{eqn:lp0}) has been recently extended in~\cite{lim2023} to include the effect of the plasma triangularity, which results into an order unity factor that multiplies $L_{p0}$. In the case considered here, this shaping factor leads only to a small positive offset of $L_{p0}$, which is neglected in the present work.
The value of $L_p$ predicted by equation~(\ref{eqn:lp0}) agrees well with the numerical results from the GBS simulations without voltage biasing. On the other hand, the theoretical prediction of $L_p$ overestimates significantly the numerical value computed from the simulations with voltage biasing, as expected since equation~(\ref{eqn:lp0}) neglects entirely the $\mathbf{E}\times\mathbf{B}$ flow shear turbulence suppression. 
Improving equation~(\ref{eqn:lp0}) to account for the voltage biasing turbulence suppression  requires one to first assess the impact of $\mathbf{E}\times\mathbf{B}$ flow shear on the quasi-linear heat flux used in \cite{giacomin2020transp} to derive equation~(\ref{eqn:lp0}), which is the aim of the following section. 

\begin{table}
    \centering
    \begin{tabular}{cccc}
    \toprule
    ${\boldsymbol{\nu_0}}$ & \multicolumn{2}{c}{\textbf{Numerical $L_p$}} & \textbf{Predicted $L_p$} \\
    & \textbf{unbiased} & \textbf{biased} & \textbf{[Eq.~(\ref{eqn:lp0})]}\\
    \midrule
    0.1 & 28 & 12 & 32\\
    1.0 & 94 & 19 & 105\\
    \bottomrule
    \end{tabular}
    \caption{Equilibrium pressure gradient length evaluated from the exponential fit in figure~\ref{fig:pressure_profiles} and the corresponding prediction from equation~(\ref{eqn:lp0}).}
    \label{tab:lp}
\end{table}

\section{Non-local linear analysis}
\label{sec:nonlocal}

The effect of the $\mathbf{E}\times\mathbf{B}$ flow shear on a quasi-linear estimate of $L_p$ is investigated here for the RBM instability by leveraging non-local linear simulations. 

\subsection{Linear model}
\label{sec:linear_model}

We consider here a reduced physical model, derived from equations~(\ref{eqn:density})-(\ref{eqn:poisson}), where only the electron pressure and the electrostatic potential is evolved, assuming $\nu j_\parallel \sim -\nabla_\parallel \phi$, neglecting the ion dynamics and considering a constant density. The pressure equation is obtained by summing equation~(\ref{eqn:density}) and equation ~(\ref{eqn:electron_temperature}), where only the $\mathbf{E}\times\mathbf{B}$ advection is retained. 
Within these approximations, the reduced physical model is given by (in normalized GBS units) 
\begin{eqnarray}
\label{eqn:pressure_red}
    \fl\qquad\frac{\partial p_e}{\partial t}  = -\rho_*^{-1} [\phi, p_e]\,,\\
    \label{eqn:phi_red}
    \fl\qquad\frac{\partial}{\partial t} \nabla \cdot (n \nabla_\perp \phi) = -\rho_*^{-1}\nabla \cdot [\phi, n\nabla_\perp \phi] + 2 C(p_e) - \frac{1}{\nu_0}\nabla_\parallel^2 \phi \,,
\end{eqnarray}
where the Reynolds stress, $\rho_*^{-1}\nabla \cdot [\phi, n\nabla_\perp \phi]$, is retained in equation~(\ref{eqn:phi_red}) to account for the $\mathbf{E}\times\mathbf{B}$ flow shear suppression mechanism. We note that this term also drives the Kelvin-Helmholtz instability at large shearing rates~\cite{Rogers2005}.

Equations~(\ref{eqn:pressure_red})-(\ref{eqn:phi_red}) are linearized assuming $p_e(x, y) = \bar{p}_e(x) + \tilde{p}_e(x)\exp(i k_y y + \gamma t)$, $\phi(x, y) = \bar{\phi}(x) + \tilde{\phi}(x)\exp(i k_y y + \gamma t)$ and $n(x, y) \simeq \bar{n}(x)$, where $\tilde{p}_e/\bar{p_e} \ll 1$ and $\phi_{1}/\phi_{0} \ll 1$, which leads to
\begin{eqnarray}
\label{eqn:pressure_lin}
    \fl \gamma \tilde{p}_e = -\rho_*^{-1} i k_y (\tilde{\phi}\partial_x \bar{p_e}    - \tilde{p}_e\partial_x \bar{\phi})\,,\\
    \label{eqn:phi_lin}
    \fl \gamma \bigl[\partial_x (\bar{n} \partial_x ) - k_y^2 \bar{n}]\tilde{\phi} =  i k_y\rho_*^{-1} \bigl[\bar{n} \partial_x\bar{\phi}\partial_x^2\tilde{\phi} - \tilde{\phi}\partial_x^2\bar{n}\partial_x\bar{\phi}-2\tilde{\phi}\partial_x\bar{n}\partial_x^2\bar{\phi} \nonumber\\- \tilde{\phi} \bar{n} \partial_x^3\bar{\phi} - k_y^2 \bar{n}\tilde{\phi}\partial_x\bar{\phi}\bigr] + 2 i k_y \tilde{p}_e + \frac{k_\parallel^2}{\nu_0}\tilde{\phi}\,,  
\end{eqnarray}
where we have replaced $\nabla_\parallel^2$ with $-k_\parallel^2$.
The linear model in equations~(\ref{eqn:pressure_lin})-(\ref{eqn:phi_lin}) retains the RBM instability as well as the effect of an equilibrium  $\mathbf{E}\times\mathbf{B}$ flow shear, which is given by the terms proportional to radial derivatives of $\bar{\phi}$. The minimal local linear physical model describing the RBM instability can be retrieved from equations~(\ref{eqn:pressure_lin})-(\ref{eqn:phi_lin}) by imposing a uniform $\bar{\phi}$ (and a uniform $\bar{n}$), leading to
\begin{eqnarray}
\label{eqn:pressure_lin_simple}
    \gamma \tilde{p}_e = -\rho_*^{-1} i k_y \tilde{\phi} \partial_x \bar{p_e} \,,\\
    \label{eqn:phi_lin_simple}
    - \gamma k_y^2 \tilde{\phi} = \frac{2}{\bar{n}} i k_y \tilde{p}_e +\frac{k_\parallel^2}{\bar{n}\nu_0} \tilde{\phi}\,.
\end{eqnarray}
From equation~(\ref{eqn:pressure_lin_simple}) and by assuming that the amplitude of $\tilde{p}_e$ saturates when the instability drive is removed from the system~\cite{Ricci2008}, i.e. $k_x \tilde{p}_e \sim \bar{p_e}/L_p$, the perpendicular turbulent heat flux is written as
\begin{equation}
\label{eqn:flux_intermediate}
q_x \sim \tilde{p}_e\partial_y\tilde{\phi} \sim \rho_* \frac{\gamma}{k_x^2}\frac{p_e}{L_p}\,.      
\end{equation}
By replacing $k_x \sim \sqrt{k_y/L_p}$~\cite{Ricci2008} in equation~(\ref{eqn:flux_intermediate}), the following quasi-linear non-local approximation of the heat flux is obtained,
\begin{equation}
\label{eqn:flux}
    q_x \sim \rho_* \bar{p_e} \frac{\gamma}{k_y}\,.
\end{equation}

A simple estimate of the pressure gradient length is derived in~\cite{ricci2013} by balancing the perpendicular and the parallel transport in the SOL, i.e. $q_x/L_p \sim q_\parallel/L_\parallel$, where $q_x$ is approximated by equation~(\ref{eqn:flux}), $q_\parallel\sim \bar{p}_ec_{s}$, and $L_\parallel \sim q\rho_*^{-1}$, leading to
\begin{equation}
\label{eqn:lp_simple}
    L_p \sim \frac{q}{c_{s}} \max\biggl(\frac{\gamma}{k_y}\biggr)\,.
\end{equation}
We highlight that $\gamma/k_y$ depends on $L_p$, i.e. equation~(\ref{eqn:lp_simple}) is an implicit equation for $L_p$, which can be solved numerically. An analytical solution of equation~(\ref{eqn:lp_simple}) can be derived by providing an analytical estimate of $\gamma$ and $k_y$. 
In~\cite{giacomin2020transp},  $\gamma$ is estimated as the maximum of the interchange instability, $\gamma \sim \sqrt{2 \bar{T}_e/(\rho_* L_p)}$, and $k_y$ is obtained by balancing the interchange driving term and the parallel current term in the reduced local model in equations~(\ref{eqn:pressure_lin_simple})-(\ref{eqn:phi_lin_simple}), leading to $k_y \simeq (\bar{n} \nu q^2 \gamma)^{-1/2}$.
On the other hand, we expect the $\mathbf{E}\times\mathbf{B}$ flow shear to directly affect $\gamma/k_y$ and, consequently, $L_p$, thus limiting the applicability of the work carried out in~\cite{giacomin2020transp}. 

\subsection{Linear simulation results}

We focus here on the results of a set of linear simulations of the physical model described by equations (\ref{eqn:pressure_lin})-(\ref{eqn:phi_lin}), with the aim of evaluating the effect of $\gamma_E$ on $\gamma/k_y$ (and therefore on $L_p$).
The linear simulations are performed with $\bar{\phi}(x) = \tanh[(x-x_0)/L_\phi]-1$, $\bar{p}_e(x) = 1 - \tanh[(x-x_0)/L_p]$, $\bar{n}(x) = 1 - \tanh[(x-x_0)/L_n]$, $x \in [100,\ 200]$, $x_0 = 150$, $N_x=200$ grid points along $x$, $k_\parallel = 1/q$, $\rho_*^{-1} = 500$, and various values of $\nu_0\in [0.05, 1.2]$, $L_p\in [10, 50]$ (with $L_n=2L_p$) and $\gamma_E \in [0, 6]$ (controlled through $L_\phi$). 

\begin{figure}
    \centering
    \subfloat[]{\includegraphics[width=0.48\linewidth]{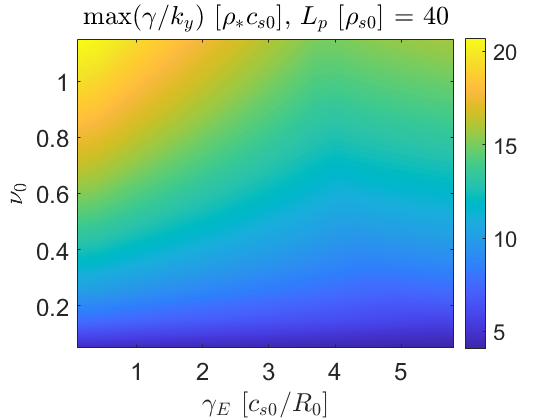}}
    \subfloat[]{\includegraphics[width=0.48\linewidth]{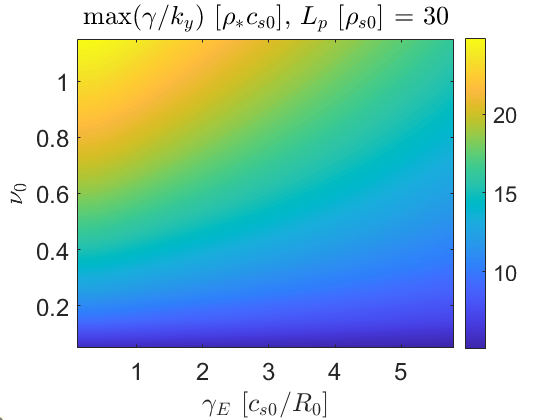}}\\
    \subfloat[]{\includegraphics[width=0.48\linewidth]{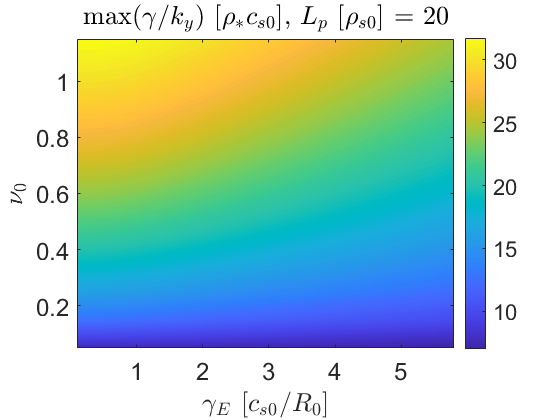}}
    \subfloat[]{\includegraphics[width=0.48\linewidth]{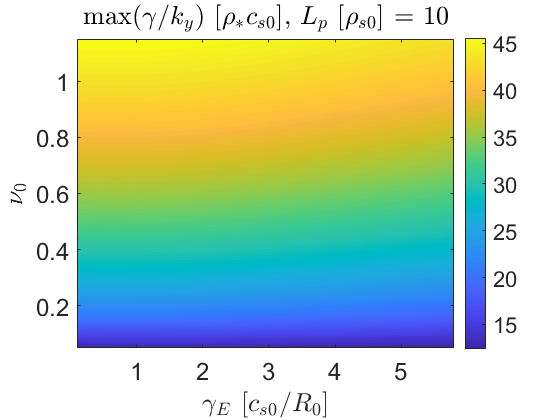}}
    \caption{Maximum value of $\gamma/k_y$ as a function of $\nu_0$ and $\gamma_E$ at four different values of $L_p$. Results from the linear solver.}
    \label{fig:gammaky}
\end{figure}

Figure~\ref{fig:gammaky} shows the maximum value of $\gamma/k_y$ as a function of $\nu_0$ and $\gamma_E$ at four different values of $L_p$. 
Starting from the case with $L_p=40$, we observe a moderate reduction of $\gamma/k_y$ as $\gamma_E$ increases, especially at $\nu_0>0.4$, while $\gamma/k_y$ is almost independent from $\gamma_E$ at very low $\nu_0$. The dependence of $\gamma/k_y$ on $\gamma_E$ reverses at large $\gamma_E$ values (see the top-right corner in figure~\ref{fig:gammaky}(a)), where the Kelvin-Helmholtz instability, which is driven by the flow shear, overcomes the RBM instability.   
The weak dependence on $\gamma_E$ at low $\nu_0$ can be explained as follows. In the RBM regime, $k_y$ is inversely proportional to $\sqrt{\nu_0}$ (explaining the increase of $\gamma/k_y$ with $\nu_0$), which leads to $k_x \sim \sqrt{k_y/L_p} \propto \nu_0^{-1/4}$. Therefore, the radial extent of RBMs increases with $\nu_0$ and radially elongated modes are more easily suppressed by the flow shear.

Looking now at different $L_p$ values, the $\gamma/k_y$ dependence on $\gamma_E$ is weaker at low $L_p$ and almost no $\gamma_E$ dependence is observed at $L_p=10$. This is a consequence of two mechanisms: (i) $k_x\propto L_p^{-1/2}$ decreases as $L_p$ increases, leading to radially narrow modes at small values of $L_p$ and hence a weaker flow shear suppression, and (ii) the flow shear suppression is expected to be negligible when $\gamma \gg \gamma_E$, where $\gamma \propto L_p^{-1/2}$ increases as $L_p$ decreases. These two combined mechanisms make the $\mathbf{E}\times\mathbf{B}$ flow shear suppression less effective at small $L_p$, i.e. at large pressure gradient.

By leveraging the results of these linear simulations, we approximate the effect of $\mathbf{E}\times\mathbf{B}$ flow shear on $L_p$ as 
\begin{equation}
\label{eqn:lp_corr}
    L_p \sim \frac{L_{p0}}{1+\alpha_k \gamma_E/\gamma}\,,
\end{equation}
where $L_{p0}$ is the equilibrium pressure gradient length when $\gamma_E=0$ and 
\begin{equation}
\label{eqn:alphak}
    \alpha_k = \frac{1-k_x/k_y}{1+k_x/k_y}
\end{equation}
is a factor that accounts for the reduction of the flow shear suppression at large $k_x$, i.e. for modes that are radially narrow. 
If $k_x\sim k_y$, the parameter $\alpha_k$ vanishes, leading to $L_p = L_{p0}$, i.e. $L_p$ is independent from $\gamma_E$, while if $k_x \ll k_y$, $\alpha_k$ tends to unity, leading to $L_p = L_{p0}/(1+\gamma_E/\gamma)$. Moreover, the $L_p$ dependence on $\gamma_E$ is negligible when $\gamma\gg \gamma_E$.  
We note that equation~(\ref{eqn:alphak}) requires $k_x < k_y$, which is satisfied in the RBM regime, but it might not hold for other microinstabilities that can develop at the tokamak edge.

\begin{figure}
    \centering
    \subfloat[]{\includegraphics[width=0.47\linewidth]{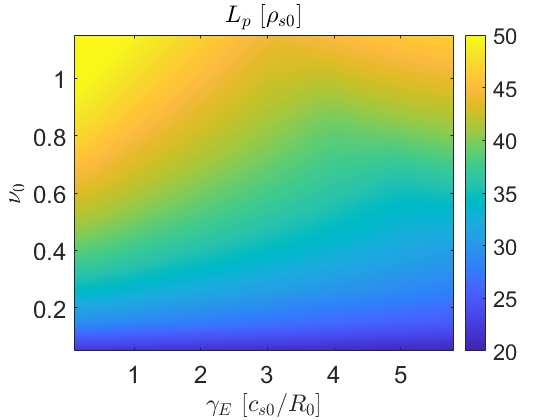}}
    \subfloat[]{\includegraphics[width=0.47\linewidth]{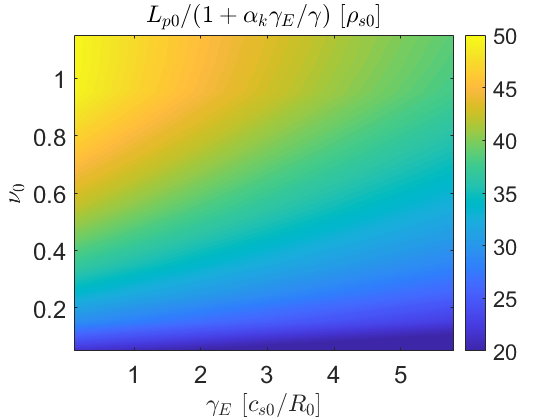}}\\
    \caption{(a) Numerical value of $L_p$ given by equation~(\ref{eqn:lp_simple}) as a function of $\nu_0$ and $\gamma_E$. (b) Theoretical estimate of $L_p$ obtained from equation~(\ref{eqn:lp_corr}) where $\gamma$ and $\alpha_k$ are evaluated at $L_{p0}$.}
    \label{fig:lp_linsolver}
\end{figure}

In principle, the factors $\gamma_E/\gamma$ and $\alpha_k$ depend on $L_p$, which makes analytical progress very challenging. 
Therefore, we evaluate $\gamma_E/\gamma$ and $\alpha_k$ at $L_{p0}$. 
Figure~\ref{fig:lp_linsolver} compares the numerical value of $L_p$ obtained from equation~(\ref{eqn:lp_simple}) using the results of the ($\nu_0$, $\gamma_E$) linear simulation scan to the value of $L_p$ obtained from equation~(\ref{eqn:lp_corr}) using solely the linear simulation results at $\gamma_E=0$ to evaluate $L_{p0}$, $\gamma$ and $\alpha_k$.
The $L_p$ approximation of equation~(\ref{eqn:lp_corr}) agrees fairly well with the numerical solution of equation~(\ref{eqn:lp_simple}), except at large $\gamma_E$ and large $\nu_0$, where $L_p$ increases with $\gamma_E$. This is expected since equation~(\ref{eqn:lp_corr}) holds only in the RBM regime, while a transition to a Kelvin-Helmholtz regime occurs at large $\gamma_E$ (the linear drive of the Kelvin-Helmholtz instability is proportional to $\gamma_E$). Excluding the Kelvin-Helmholtz instability, which occurs at shearing rates much larger than those achievable in RFX-mod, we note a modest decrease of $L_p$ as $\gamma_E$ increases at intermediate and large $\nu_0$ values, while a negligible dependence of $L_p$ on $\gamma_E$ is observed at low $\nu_0$, which is consistent with figure~\ref{fig:gammaky}.

Before concluding this section, we note that equation~(\ref{eqn:lp_corr}) extends the analytical estimate of $L_p$ derived in~\cite{giacomin2020transp} in the absence of $\mathbf{E}\times\mathbf{B}$ flow shear to cases of non-negligible $\mathbf{E}\times\mathbf{B}$ flow shear, such as the ones considered in section~\ref{sec:gbs}. An analytical estimate of the edge equilibrium pressure gradient length in the presence of non-negligible flow shear is derived in the next section starting from equation~(\ref{eqn:lp_corr}).

\section{Equilibrium pressure gradient in the presence of voltage biasing}
\label{sec:theory}

In this section, we provide an analytical estimate of the factors $\gamma_E/\gamma$ and $\alpha_k$, which account for the $\mathbf{E}\times\mathbf{B}$ flow shear suppression in equation~(\ref{eqn:lp_corr}).
The analytical estimates are first derived in terms of GBS normalized quantities, allowing for a direct comparison with GBS simulation results, and then written in terms of engineering parameters for an easier comparison to experiments. 
The consequences of the improved $L_p$ estimate on the density limit scaling law derived in~\cite{giacomin2022density} are also discussed. 

\subsection{Theoretical derivation and comparison to GBS results}
\label{sec:derivation}

An analytical estimate of $\gamma_E/\gamma \sim \gamma_E\sqrt{\rho_* L_{p0}/(2 T_e)}$ is obtained by substituting $L_{p0}$ from equation~(\ref{eqn:lp0}), which leads to
\begin{equation}
\label{eqn:gammae}
    \frac{1}{\gamma} \sim \frac{\nu^{1/3}n}{2^{2/3}}\biggl(\frac{\rho_* q L_\chi T_e}{S_p}\biggr)^{2/3}\,,
\end{equation}
where, here and in the following, $n$ and $T_e$ denote the equilibrium density and equilibrium electron temperature near the separatrix, respectively, with the overline dropped for clarity purposes.

The factor $\alpha_k$ depends on the ratio $k_x/k_y$. By estimating $k_y$ as the poloidal wave-vector corresponding to the maximum of $\gamma/k_y$, i.e. $k_y\sim (n\nu q^2\gamma)^{-1/2}$, and by considering  $k_x \sim \sqrt{k_y/L_{p0}}$, we obtain 
\begin{equation}
\label{eqn:kxky_first}
\frac{k_x}{k_y} \sim \frac{(n\nu q^2)^{1/4}}{L_{p0}^{1/4}}\biggl(\frac{2 T_e}{\rho_* L_{p0}}\biggr)^{1/8}\,.
\end{equation}
Substituting $L_{p0}$ into equation~(\ref{eqn:kxky_first}) leads to
\begin{equation}
    \label{eqn:kxoky}
    \frac{k_x}{k_y}\sim 2^{1/3}\biggl(\frac{S_p^5}{\rho_*^2q^2L_\chi^5 \nu T_e^8n^6}\biggr)^{1/6}\,.
\end{equation}
The factor $\alpha_k$ is then computed from equation~(\ref{eqn:kxoky}),
\begin{equation}
\label{eqn:alphak_full}
    \alpha_k \sim \frac{1 - 2^{1/3}S_p^{5/6}(\rho_*^2q^2L_\chi^5 \nu T_e^8n^6)^{-1/6}}{1+ 2^{1/3}S_p^{5/6}(\rho_*^2q^2L_\chi^5 \nu T_e^8n^6)^{-1/6}}\,,
\end{equation}
which holds only when $k_x<k_y$. While this condition is in general satisfied in the RBM regime, other turbulent transport regimes may have $k_x>k_y$, thus preventing a straightforward generalization of equation~(\ref{eqn:alphak_full}), which is specific to the RBM regime.  

Equations~(\ref{eqn:gammae})~and~(\ref{eqn:alphak_full}) can be easily evaluated for the GBS simulations performed in section~\ref{sec:gbs}. 
For the high-$\nu_0$ GBS simulation, we obtain $\gamma_E/\gamma \simeq 8.1$, $k_x/k_y \simeq 0.23$ and $\alpha_k \simeq 0.62$, leading to $L_p \simeq 0.16 L_{p0}\simeq 17$, which is close to the numerical $L_p$ value computed from the high-$\nu_0$ GBS simulation with edge voltage biasing ($L_p=19$ from table~\ref{tab:lp}). 
For the low-$\nu_0$ GBS simulation, we obtain $\gamma_E/\gamma \simeq 4.9$, $k_x/k_y \simeq 0.25$ and $\alpha_k \simeq 0.6$, leading to $L_p \simeq 0.3 L_{p0}\simeq 10$, which agrees with  the numerical value computed from the low-$\nu_0$ GBS simulation with edge voltage biasing ($L_p=12$ from table~\ref{tab:lp}).

The suppression factor $(1+\alpha_k\gamma_E/\gamma)^{-1}$ is almost a factor of two larger at $\nu_0=1.0$ than at $\nu_0=0.1$,  although $\gamma_E$ achieves larger values at $\nu_0=1.0$ (see figure~\ref{fig:shear_profiles}). This is in agreement with the GBS simulation results of section~\ref{sec:gbs}, where a stronger $\mathbf{E}\times\mathbf{B}$ flow shear suppression is observed in the simulation with larger $\nu_0$ (and larger $L_p$). 
In conclusion, equation~(\ref{eqn:lp_corr}) with $\gamma_E/\gamma$ given by equation~(\ref{eqn:gammae}) and $\alpha_k$ given by equation~(\ref{eqn:alphak}) extends the applicability of the $L_p$ theoretical scaling law of~\cite{giacomin2020transp} including  turbulence suppression from the $\mathbf{E}\times\mathbf{B}$ flow shear.    

In~\cite{giacomin2021}, the electron temperature dependence of $L_{p0}$ has been explicitly written in terms of $S_p$ by considering a balance between the total heating source, $S_p$, and the parallel losses in the SOL, $q_\parallel\simeq p_e c_s$, leading to $T_e\simeq (5 S_p/(4nL_{p0}))^{2/3}$ or, using the analytical estimate of $L_{p0}$ given in equation~(\ref{eqn:lp0}), to
\begin{equation}
\label{eqn:te_conv}
    T_e \sim \frac{5^{6/17}}{2^{10/17}}\rho_*^{-2/17}S_p^{14/17}L_\chi^{-8/17}n^{-18/17}q^{-8/17}\nu_0^{-4/17}\,.
\end{equation}
For consistency with~\cite{giacomin2021}, we substitute the electron temperature estimate given in equation~(\ref{eqn:te_conv}) into equations~(\ref{eqn:gammae})~and~(\ref{eqn:kxoky}), which leads to
\begin{equation}
\label{eqn:gammae_notemp}
    \frac{1}{\gamma} \sim \frac{5^{1/17}}{2^{13/17}}  L_\chi^{10/17}n^{14/17}q^{10/17}\nu_0^{5/17}\rho_*^{11/17}S_p^{-9/17}
\end{equation}
and
\begin{equation}
\label{eqn:kxoky_notemp}
    \frac{k_x}{k_y} \sim \frac{2^{33/34}}{5^{13/34}}\rho_*^{-7/34}n^{5/34}q^{3/17}\nu_0^{3/34}L_\chi^{-11/34}S_p^{-1/17}\,.
\end{equation}
We note that equations~(\ref{eqn:gammae})~and~(\ref{eqn:gammae_notemp}) share a very similar dependence on $n$ and $S_p$, while the dependence of $k_x/k_y$ on $n$ and $S_p$ in equation~(\ref{eqn:kxoky_notemp}) is modified with respect to the one in equation~(\ref{eqn:kxoky}) and the suppression factor $\alpha_k$ computed from equation~(\ref{eqn:kxoky_notemp}) is found to depend mainly on geometrical parameters.

\subsection{Suppression factors in terms of engineering parameters}
\label{sec:eng_par}

For an easier comparison with the experiments, we write here the $\mathbf{E}\times\mathbf{B}$ flow shear suppression factors in terms of engineering parameters. 
By restoring the physical units ($n_\mathrm{\scriptscriptstyle{GBS}} =  n/n_0$, $T_{e,\mathrm{\scriptscriptstyle{GBS}}} = T_e / T_{e0}$, $L_{\chi,\mathrm{\scriptscriptstyle{GBS}}} = L_\chi / \rho_{s0}$ and $ S_{p,\mathrm{\scriptscriptstyle{GBS}}} = S_p \Omega_{ci} / (n_0 T_{e0} c_{s0}^2)$) and substituting $\nu_0$ with its definition given in equation~(\ref{eqn:conductivity}), $1/\gamma$ becomes
\begin{equation}
    \label{eqn:gammae_phys}
    \fl\qquad\frac{1}{\gamma} \sim 2^{-1/3} q^{2/3}\frac{R_0}{c_s}\biggl(\frac{1}{1.96}\frac{m_e}{m_i}\frac{R_0}{c_s\tau_e}\biggr)^{1/3}\biggl(\frac{L_\chi}{R_0}\biggr)^{2/3}\biggl(\frac{n T_e c_s^2}{S_p\Omega_{ci}}\biggr)^{2/3}\,.
\end{equation}
By replacing $S_p = P_\mathrm{SOL}/(2\pi R_0)$, $L_\chi = \sqrt{2(1+\kappa^2)}\pi a$ and $\tau_e$ (see equation~(\ref{eqn:resistivity})) in equation~(\ref{eqn:gammae_phys}), we obtain
\begin{equation}
\label{eqn:gammae_eng}
    \fl\qquad\frac{1}{\gamma} \simeq 1.8\cdot 10^{-7} A^{1/3} a^{2/3} (1+\kappa^2)^{1/3}  q^{2/3} R_0^{4/3} B_0^{-2/3} n T_e^{1/6} P_\mathrm{SOL}^{-2/3}\,,
\end{equation}
where $1/\gamma$ is in units of $\mathrm{s}$, $A$ is the isotope mass number, $R_0$ and $a$ are the minor and major radii in m, $\kappa$ is the plasma elongation at the last-closed flux surface, $B_0$ is the toroidal magnetic field in T, $n$ is the separatrix density in $10^{19}\mathrm{m}^{-3}$, $T_e$ is the separatrix electron temperature in eV, and $P_\mathrm{SOL}$ is the power crossing the separatrix in MW. The factor $\gamma_E/\gamma$ is readily obtained by multiplying equation~(\ref{eqn:gammae_eng}) by the $\mathbf{E}\times\mathbf{B}$ shearing rate in units of $\mathrm{s}^{-1}$.

Similarly, the ratio $k_x/k_y$ in physical units reads 
\begin{equation}
    \fl\quad\frac{k_x}{k_y} \sim 2^{1/3}q^{-1/3}\biggl(\frac{S_p\Omega_{ci}}{nT_e c_s^2}\biggr)^{5/6}\biggl(\frac{L_\chi}{\rho_s}\biggr)^{-5/6}\biggl(\frac{1}{1.96}\frac{m_e}{m_i}\frac{R_0}{c_s}\frac{1}{\tau_e}\biggr)^{-1/6}\biggl(\frac{R_0}{\rho_s}\biggr)^{1/3}\,,
\end{equation}
which, in terms of engineering parameters, leads to
\begin{equation}
\label{eqn:kxoky_eng}
    \fl\quad\frac{k_x}{k_y} \simeq 36 A^{1/3} a^{-5/6} (1+\kappa^2)^{-5/12}R_0^{-2/3}q^{-1/3}B_0^{1/3}P_\mathrm{SOL}^{5/6}n^{-1}T_e^{-13/12}\,,
\end{equation}
where all the quantities are evaluated with the same units of equation~(\ref{eqn:gammae_eng}). The factor $\alpha_k$ in terms of engineering parameters can be readily evaluated from equations~(\ref{eqn:alphak})~and~(\ref{eqn:kxoky_eng}).

We can now easily estimate $\gamma_E/\gamma$ and $\alpha_k$ for the RFX-mod reference discharge, where $\gamma_E\simeq 10^6 s^{-1}$, $n\simeq 2\cdot 10^{18}$~m$^{-3}$, $T_e\simeq20$~eV, $B_0\simeq 0.55$~T, $q\simeq 3.5$, $R_0\simeq 2$~m, $a\simeq0.5$~m, $A=2$, and $P_\mathrm{SOL}\simeq80$~kW. This leads to $\gamma_E/\gamma \simeq 2.8$, $k_x/k_y\simeq 0.48$, $\alpha_k \simeq 0.34$, and $(1+\alpha_k\gamma_E/\gamma)^{-1}\simeq 0.5$.
The theoretical scaling predicts, for the RFX-mod reference discharge, a factor two larger pressure gradient at the separatrix  in the biased phase than in the unbiased one. This pressure gradient increase is consistent with the L-H transition induced by the biasing electrode in RFX-mod experiments~\cite{spolaore2017,grenfell2020}.  
The theoretical scaling laws of the suppression factors given in equations~(\ref{eqn:gammae_eng})~and~(\ref{eqn:kxoky_eng}) seem to describe fairly well the experimental pressure profile steeping observed in RFX-mod experiments when the tokamak edge is negatively biased.

For completeness, we also report here equations~(\ref{eqn:gammae_notemp})~and~(\ref{eqn:kxoky_notemp}) in terms of engineering parameters,
\begin{eqnarray}
\label{eqn:geog_eng_notemp}
    \fl\quad\frac{1}{\gamma} &\sim& 4\cdot 10^{-7} A^{13/34} a^{10/17}R_0^{20/17}(1+\kappa^2)^{5/17}n^{14/17}q^{10/17}B_0^{-10/17}P_\mathrm{SOL}^{-9/17}\,,\\
    \label{eqn:kxoky_eng_notemp}
    \fl\quad\frac{k_x}{k_y} &\sim& 0.2 A^{1/68} a^{-11/34}(1+\kappa^2)^{-11/68}n^{5/34}q^{3/17}R_0^{6/17}B_0^{-3/17}P_\mathrm{SOL}^{-1/17}\,.
\end{eqnarray}
As expected, $\gamma_E/\gamma$ and $k_x/k_y$ evaluated from equations~(\ref{eqn:geog_eng_notemp})~and~(\ref{eqn:kxoky_eng_notemp}) on the RFX-mod reference discharge return values similar to those obtained from equations~(\ref{eqn:gammae_eng})~and~(\ref{eqn:kxoky_eng}). 

\subsection{Consequences on the density limit}
\label{sec:density_limit}

The density limit theory proposed in~\cite{giacomin2022density} describes the transition to very large turbulent fluxes observed in GBS simulations at high density, but it is unable to predict the result of the high-$\nu_0$ GBS biased simulation presented in section~\ref{sec:gbs}. In fact, the work of~\cite{giacomin2022density} considers only the self-generated $\mathbf{E}\times\mathbf{B}$ flow shear, which is very weak at high density.     
We extend here the density limit scaling of~\cite{giacomin2022density} by considering the improved $L_p$ estimate that accounts for  $\mathbf{E}\times\mathbf{B}$ flow shear turbulence suppression.
The criterion for the crossing of the density limit defined in~ \cite{giacomin2022density}, i.e. $L_{p0}\sim a$, becomes therefore $L_{p0}\sim a(1+\alpha_k\gamma_E/\gamma)$.

We note that both $\gamma_E/\gamma$ and $\alpha_k$ depend on density. However, making analytical progress while retaining the density dependence in $\alpha_k$ is particularly challenging. In addition, as discussed in section~\ref{sec:derivation}, $k_x/k_y$ in the RBM regime considered here turns out to depend weakly on density. Therefore, the parameter $\alpha_k$ is approximated here as an additional geometrical factor that provides an order unity correction to $\gamma_E/\gamma$. Within this approximation, the condition for density limit crossing is  written as 
\begin{equation}
\label{eqn:dl_eqn}
    n_\mathrm{DL}^2 - 2 \alpha_{\gamma_E} n_\mathrm{DL0} n_\mathrm{DL} - n_\mathrm{DL0}^2 =0\,,
\end{equation}
where 
\begin{equation}
     n_\mathrm{DL0} \sim 2^{1/6} \pi^{-2/3} \rho_*^{-1/6} \nu^{-1/3} q^{-2/3}(1+\kappa^2)^{-1/3} a^{-1/6} T_e^{-7/6} S_p^{2/3}
\end{equation}
is the maximum achievable edge density derived in~\cite{giacomin2022density} in the absence of $\mathbf{E}\times\mathbf{B}$ flow shear and 
\begin{equation}
\label{eqn:alpha_ge}
    \alpha_{\gamma_E} \sim 2^{-3/2}\alpha_k\gamma_E T_e^{-1/2}\rho_*^{1/2}a^{1/2}
\end{equation}
accounts for the $\mathbf{E}\times\mathbf{B}$ flow shear turbulence suppression, such that $\alpha_{\gamma_E}=0$ leads to $n_\mathrm{DL}=n_\mathrm{DL0}$.
A scaling law for $n_\mathrm{DL}$ is obtained from equation~(\ref{eqn:dl_eqn}), i.e.
\begin{equation}
\label{eqn:dl}
    n_\mathrm{DL} = n_\mathrm{DL0} \Bigl( \alpha_{\gamma_E} + \sqrt{1 + \alpha_{\gamma_E}^2}\Bigr)\,.
\end{equation}
As expected, the maximum achievable edge density predicted by equation~(\ref{eqn:dl}) increases with the $\mathbf{E}\times\mathbf{B}$ flow shear suppression factor $\alpha_{\gamma_E}$, which depends linearly on $\gamma_E$.

By following~\cite{giacomin2022density}, the separatrix electron temperature in equation~(\ref{eqn:alpha_ge}) is replaced by the two-point model prediction~\cite{stangeby2000}, 
\begin{equation}
\label{eqn:te}
T_e \simeq \biggl(\frac{7}{2} \frac{S_p}{\chi_{\parallel e 0}} \frac{q}{a \rho_*} \frac{L_\parallel}{L_{p0}}\biggr)^{2/7}\,,   
\end{equation}
where $L_\parallel$ is the SOL parallel connection length, $\chi_{\parallel e 0}$ is defined in equation~(\ref{eqn:chie}) and $L_p$ is approximated with $L_{p0}$. 
This leads to  
\begin{equation}
\label{eqn:alpha_ge_notemp}
    \alpha_{\gamma_E} \sim \frac{\alpha_k}{2^{19/14}7^{1/7}}  \gamma_E \rho_*^{9/14} a^{11/14} S_p^{-1/7}\chi_{\parallel e 0}^{1/7} q^{-1/7} L_\parallel^{-1/7}\,,
\end{equation}
which depends linearly on $\gamma_E$ and weakly on the parallel connection length, on the heating source and on the parallel conductivity (the exponent is $1/7\simeq 0.14$).

In order to allow for an easy comparison with experiments, the turbulence suppression factor $\alpha_{\gamma_E}$ is written in terms of engineering parameters ($n_\mathrm{\scriptscriptstyle{GBS}} =  n/n_0$, $T_{e,\mathrm{\scriptscriptstyle{GBS}}} = T_e / T_{e0}$, $S_p = P_\mathrm{SOL}/(2\pi R_0)$ and $\chi_{\parallel e 0}$ given by equation~(\ref{eqn:chie})), leading to 
\begin{equation}
    \label{eqn:alpha_ge_phys}
    \alpha_{\gamma_E} \simeq 1.6\cdot 10^{-5} \alpha_k \gamma_E A^{1/2} R^{1/2} a^{11/14} P_\mathrm{SOL}^{-1/7}q^{-1/7}L_\parallel^{-1/7}\,,
\end{equation}
with $\gamma_E$ in units of $s^{-1}$, $R$ and $a$ in m, $P_\mathrm{SOL}$ in MW, and $L_\parallel$ in m. 
We note that $\alpha_{\gamma_E}$ increases with both $R_0$ and $a$, suggesting a larger $\mathbf{E}\times\mathbf{B}$ flow shear turbulence suppression in larger tokamaks for the same $\gamma_E$ value. 
The impact of the voltage biasing on the density limit of a RFX-mod plasma with voltage biasing is readily evaluated from equation~(\ref{eqn:alpha_ge_phys}). By considering $q\simeq 3.5$, $R_0\simeq 2$~m, $L_\parallel \simeq q R \simeq 7$~m, $a\simeq 0.5$~m, $A=2$, $P_\mathrm{SOL}\simeq80$~kW, $\alpha_k\simeq 0.34$ (evaluated in section~\ref{sec:eng_par}) and $\gamma_E\simeq 10^5 s^{-1}$\footnote{We consider here a rather conservative value of $\gamma_E$, which is smaller than the experimental value measured in the low density RFX-mod discharge \#39136. In fact, we assume that experimentally achievable values of $\gamma_E$ through voltage biasing scale inversely with density as the drained current.}, we obtain $\alpha_{\gamma_E} \simeq 0.6$. Substituting this value into equation~(\ref{eqn:dl}) yields to $n_\mathrm{DL}/n_\mathrm{DL0} \simeq 1.8$.
Equation~(\ref{eqn:alpha_ge_phys}) suggests that the density limit could be increased by almost a factor of two in RFX-mod through edge voltage biasing.
However, dedicated future experiments in RFX-mod2 are needed to confirm the validity of the predictions returned by equation~(\ref{eqn:alpha_ge_phys}).

\section{Conclusions}
\label{sec:conclusions}

Edge voltage biasing provides a direct mechanism to drive $\mathbf{E}\times\mathbf{B}$ flow shear, allowing for investigations of its effects on turbulent transport in isolation from other mechanisms.
The effect of the edge voltage biasing on turbulent transport is analyzed in the present work by means of three-dimensional flux-driven two-fluid global turbulence simulations carried out with the GBS code, which is extended here to include a biasing electrode. The simulations have been performed considering the  RFX-mod diverted plasma of the discharge \#39136~\cite{grenfell2020}, where the biasing electrode has been used to generate a significant level of $\mathbf{E}\times\mathbf{B}$ flow shear that enables the access to the high-confinement regime.
A total of four GBS simulations are considered: two simulations at low reference density (low $\nu_0$), near the experimental value of the separatrix density in \#39136, and two simulations at high reference density (high $\nu_0$), in proximity of the crossing of the density limit. In both cases, the simulations have been performed with and without voltage biasing, where the voltage biasing value is chosen to match the one achieved in the reference RFX-mod discharge, both in the low-$\nu_0$ and high-$\nu_0$ simulations.
The current drained by the electrode in the low-$\nu_0$ simulation  is within typical current values drained by the electrode in RFX-mod experiments, while it is approximately an order of magnitude larger in the high-$\nu_0$ simulation.
In order to reduce the computational cost of these simulations, the size of the simulation domain is approximately half of the RFX-mod size.
All the GBS simulations considered in this work are in the RBM regime described in~\cite{giacomin2022turbulent}.

At both values of $\nu_0$, GBS simulations show a strong turbulent transport reduction when the tokamak edge is biased, which is caused by the large $\mathbf{E}\times\mathbf{B}$ flow shear that forms across the separatrix.
In particular, the collapse of the pressure gradient, observed in the high-density unbiased simulation and corresponding to the crossing of the density limit, is completely avoided in the biased simulation, thus suggesting an important role played by the $\mathbf{E}\times\mathbf{B}$ flow shear turbulence suppression in the density limit physics. 

The heat and particle transport reduction due to $\mathbf{E}\times\mathbf{B}$ flow shear turbulence suppression leads to the formation of a steep pressure gradient across the separatrix, which is consistent with a pedestal in the low density simulation, thus reproducing qualitatively the H-mode access observed in the RFX-mod discharge \#39136 when the tokamak edge is biased. 
The value of the equilibrium pressure gradient length evaluated from the GBS simulations is found to agree well with the prediction returned by the analytical $L_p$ scaling of~\cite{giacomin2020transp} in the unbiased cases, while the numerical $L_p$ is significantly smaller than the theoretical prediction in the biased cases, especially at large density where the theoretical $L_p$ is approximately a factor of five larger than the numerical one. 

In order to improve the theoretical $L_p$ scaling of~\cite{giacomin2020transp} and account for the $\mathbf{E}\times\mathbf{B}$ flow shear turbulence suppression, a non-local linear analysis is carried out to investigate the impact of $\gamma_E$ on a quasi-linear estimate of $L_p$ in the RBM regime. By leveraging the results of a set of linear simulations performed at various values of $\nu_0$ and $\gamma_E$, an improved $L_p$ scaling is identified, $L_p \sim L_{p0}/(1+\alpha_k \gamma_E/\gamma)$, where $L_{p0}$ is the theoretical scaling in the absence of $\mathbf{E}\times\mathbf{B}$ flow shear, while $(1+\alpha_k \gamma_E/\gamma)^{-1}$  is a suppression factor that includes the $\mathbf{E}\times\mathbf{B}$ velocity shearing rate. An analytical estimate of the factors $\gamma_E/\gamma$ and $\alpha_k$ is provided in equations~(\ref{eqn:gammae})~and~(\ref{eqn:alphak_full}), respectively. 
The improved $L_p$ scaling is found to reproduce well the pressure gradient steepening observed in the GBS simulations with edge voltage biasing.

In order to allow for an easier comparison with experiments, the analytical scaling laws of the $\mathbf{E}\times\mathbf{B}$ suppression factors are provided in terms of engineering parameters (see equations~(\ref{eqn:geog_eng_notemp})~and~(\ref{eqn:kxoky_eng_notemp})). 
Evaluating the suppression factors for the RFX-mod reference discharge returns a pressure gradient that is a factor of two larger in the biased phase compared to the unbiased phase. This result is consistent with the pressure profile steepening associated with the L-H transition induced by the edge voltage biasing in the RFX-mod discharge \#39136. However, we highlight that the theoretical scaling derived in this work is restricted to the RBM regime, and its generalization to other turbulent transport regimes would require dedicated investigations, which are outside the scope of the present work.

The consequences of the improved $L_p$ estimate on the density limit scaling law derived in~\cite{giacomin2022density} are also analyzed, and the condition for the density limit crossing considered in~\cite{giacomin2022density} is modified to $L_{p0}\simeq a(1+\alpha_k\gamma_E/\gamma)$, thus accounting for the $\mathbf{E}\times\mathbf{B}$ flow shear turbulence suppression induced by edge voltage biasing. An improved density limit scaling law is derived and written in terms of engineering parameters for an easy evaluation (see equation~(\ref{eqn:dl})). By considering a rather conservative shearing rate of $\gamma_E=10^5$~s$^{-1}$, the scaling law derived in this work predicts that the maximum achievable edge density in the RFX-mod reference discharge could be increased by a factor of two by leveraging edge voltage biasing.
Future RFX-mod2 experiments are needed to investigate whether relevant shearing rate values can be achieved with the biasing electrode at large density and to validate the proposed density limit scaling law in the presence of $\mathbf{E}\times\mathbf{B}$ flow shear.

We conclude by remarking that the results derived herein consider the $\mathbf{E}\times\mathbf{B}$ velocity shear to be an external actuator that can be directly modified through the voltage biasing, mimicking RFX-mod experiments. On the other hand, the scaling laws derived here are expected to hold even when the $\mathbf{E}\times\mathbf{B}$ velocity shear is generated by other mechanisms, since $L_p$ retains the dependence on additional parameters, such as the power crossing the separatrix and the safety factor, which may vary when extra rotation is driven. 
As a future work, we will carry out an extensive validation of a full-size RFX-mod2 turbulence simulation with edge voltage biasing, considering, in particular, a high-density discharge.

\section*{Acknowledgments}

The authors would like to thank Paolo Ricci and the GBS team for very useful discussions, Domenico Abate for providing the RFX-mod magnetic equilibrium, and Italo Predebon for a useful feedback. 
This work has been carried out within the framework of Italian National Recovery and Resilience Plan (NRRP), funded by the European Union - NextGenerationEU (Mission 4, Component 2, Investment 3.1 - Area ESFRI Energy - Call for tender No. 3264 of 28-12-2021 of Italian University and Research Ministry (MUR), Project ID IR0000007, MUR Concession Decree No. 243 in date 04/08/2022, CUP B53C22003070006, ``NEFERTARI – New Equipment for Fusion Experimental Research and Technological Advancements with Rfx Infrastructure'').
Views and opinions expressed are however those of the author(s) only and do not necessarily reflect those of the European Union or the European Commission. Neither the European Union nor the European Commission can be held responsible for them.
The GBS simulations have been carried out on the CINECA Marconi HPC facility under the EUROfusion HPC project ``FUA38\_BTIHM''.

\appendix
\section{Comparison between the numerical and experimental flow shear radial profile}
\label{sec:comparison}

A detailed validation of the GBS simulations with (and without) voltage biasing of a RFX-mod plasma would require performing full-size RFX-mod (or RFX-mod2) simulations as well as a dedicated set of experimental discharges where turbulence properties and equilibrium profiles are accurately measured in the plasma boundary, similarly to the validation carried out recently on TCV~\cite{oliveira2022}. This detailed validation, which is outside the aim of the present work, will be performed in future. 
Herein, we show that, despite the reduced size of the simulation domain, the radial profiles of the electrostatic potential and of the $\mathbf{E}\times\mathbf{B}$ flow shear from the simulations with $\nu_0=0.1$, with and without voltage biasing, agree reasonably well with those measured at the outer mid-plane in the RFX-mod discharge \#39136. This comparison provides a first validation of the biasing electrode implementation in GBS. 

\begin{figure}
    \centering
    \subfloat[Floating potential.]{\includegraphics[width=0.48\textwidth]{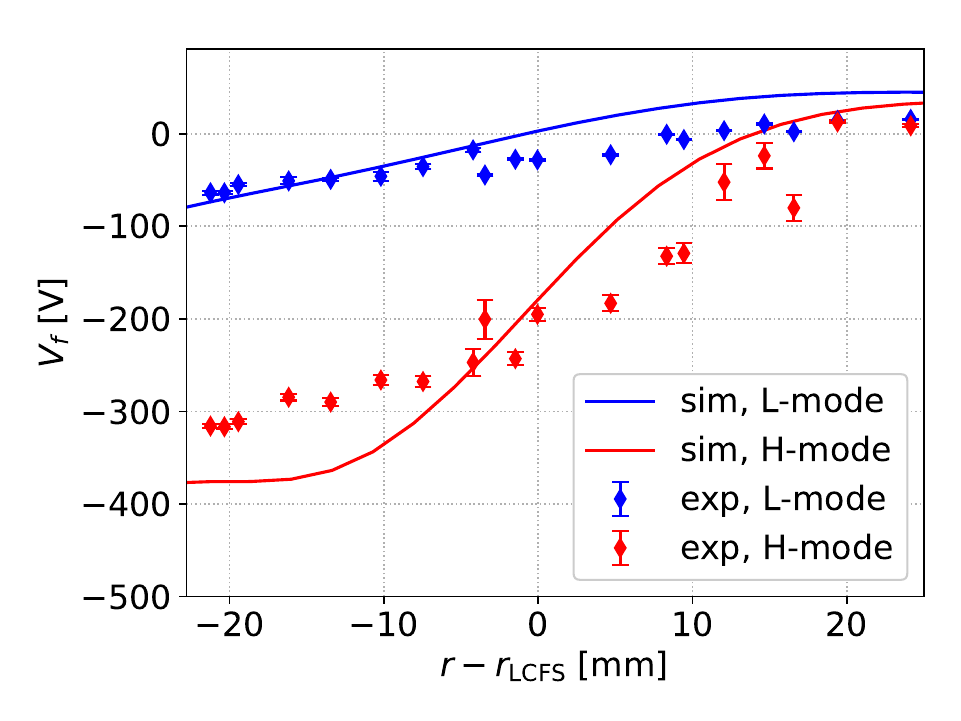}}
    \subfloat[$\mathbf{E}\times\mathbf{B}$ flow shear.]{\includegraphics[width=0.48\textwidth]{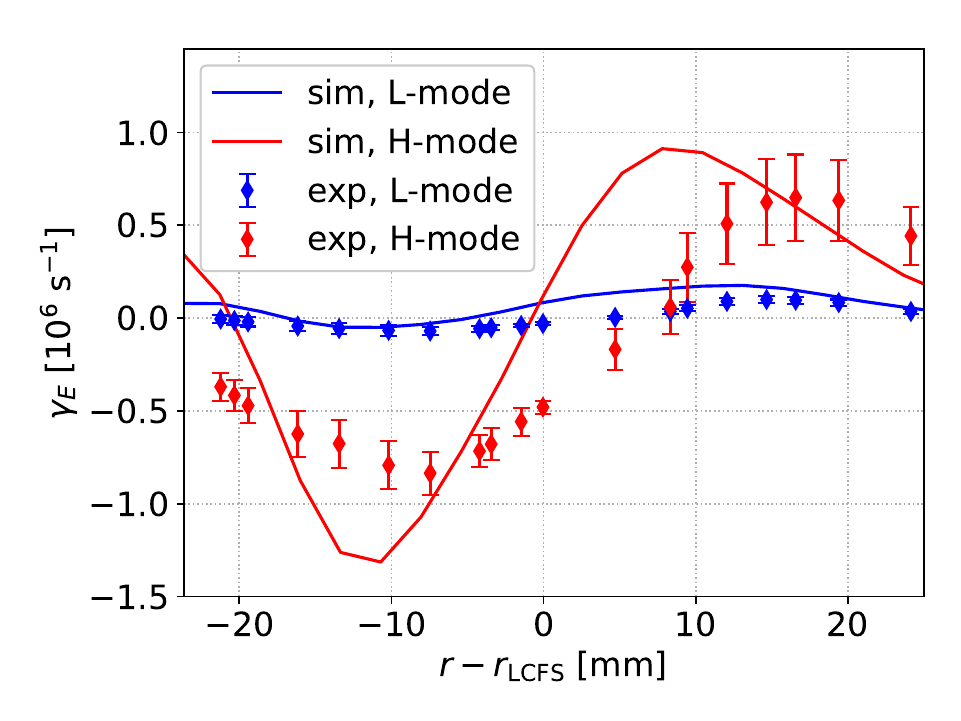}}
    \caption{Comparison between the experimental (markers) and the numerical (solid line) floating potential (a) and $\mathbf{E}\times\mathbf{B}$ flow shear (b) outboard mid-plane radial profiles in the L-mode (blue) and H-mode (red) phase. The numerical profiles are evaluated from the GBS simulations at $\nu_0=0.1$. The variable $r_\mathrm{LCFS}$ denotes the position of the separatrix.}
    \label{fig:exp_comp}
\end{figure}

Figure~\ref{fig:exp_comp} compares the outboard mid-plane radial profile of the floating potential from the GBS simulations with $\nu_0=0.1$, computed as $V_f = \phi - \Lambda T_e$ and converted into physical units, to the experimental one, obtained from the U-probe data (see~\cite{spolaore2017} for details on the U-probe).
Similarly to~\cite{grenfell2020}, the floating potential is measured in the discharge \#39136 between 470~ms and 570~ms for the L-mode phase and between 620~ms and 720~ms for the H-mode phase. The experimental $\mathbf{E}\times\mathbf{B}$ flow shear radial profile is computed as $\gamma_E \sim -\partial_{rr}^2V_f/B$ from an interpolated curve of the floating potential data after applying a third-order Savitsky--Golay filter to smooth the interpolated curve. 
In order to evaluate the experimental uncertainty on the flow shear radial profile, the time window of 100~ms, where the floating potential is measured, is divided in ten sub-intervals, and a radial profile of $\gamma_E(r)$ is computed for every sub-intervals. The experimental uncertainty on the averaged $\gamma_E(r)$ radial profile is finally obtained as standard deviation.
We expect this method to account, at least partially, for the uncertainty deriving from smoothing the experimental floating potential, which is necessary to be able to evaluate its second-order radial derivative. 

In the L-mode phase, the floating potential from the unbiased GBS simulation with $\nu_0=0.1$ agrees well with the measurements inside the separatrix ($r<r_\mathrm{LCFS}$), while GBS overestimates slightly the floating potential in the far SOL ($r-r_\mathrm{LCFS}>10$~mm).  Moreover, the numerical $\mathbf{E}\times\mathbf{B}$ flow shear profile agrees qualitatively and quantitatively with the experimental one, except in the region $0 < r-r_\mathrm{LCFS}<10$~mm, where the flow shear values from the unbiased simulation are slightly larger than the experimental ones.
In the H-mode phase, the strong decrease of the floating potential induced by the biasing electrode in the experiment is reproduced relatively well in the GBS simulation with the electrode, although a small quantitative disagreement is observed in the region $-20\,\mathrm{mm}<r-r_\mathrm{LCFS}< -10\,\mathrm{mm}$. GBS also reproduces reasonably well, both qualitatively and quantitatively, the experimental flow shear profile, except for the peak in the SOL, which is closer to the separatrix in the simulation than in the experiment, although the experimental value of the flow shear in this region is affected by a large uncertainty and the position of the peak might not be accurately measured. 
Overall, the flow shear values evaluated from the GBS simulations in the region across the separatrix are comparable to those evaluated from the floating potential measurements performed in the RFX-mod reference discharge, both in L-mode and in H-mode.

\section*{References}
\bibliographystyle{unsrt}
\bibliography{library}

\end{document}